\documentclass[pra,aps,twocolumn, nobalancelastpage, superscriptaddress]{revtex4-1}
\usepackage[utf8]{inputenc}
\usepackage[utf8]{inputenc}
\usepackage{bbold}
\usepackage{geometry}
\usepackage{float}
\usepackage{graphicx}

\graphicspath{ {images/} }
\usepackage{titlesec} 
\usepackage{braket}
\usepackage[normalem]{ulem}
\usepackage[T1]{fontenc}
\usepackage{amsmath,amsfonts,amssymb}
\usepackage{diagbox}
\usepackage{amsthm}
\usepackage[]{natbib}
\usepackage{comment}

\newcommand{\be}{\begin{equation}}
\newcommand{\ee}{\end{equation}}
\newcommand{\ba}{\begin{aligned}}
\newcommand{\ea}{\end{aligned}}

\newcommand{\bc}{\begin{center}}
\newcommand{\ec}{\end{center}}
\newcommand{\beq}{\begin{equation}}
\newcommand{\eeq}{\end{equation}}
\newcommand{\beqq}{\begin{equation*}}
\newcommand{\eeqq}{\end{equation*}}
\newcommand{\beqa}{\begin{align}}
\newcommand{\eeqa}{\end{align}}
\newcommand{\barr}{\begin{array}}
\newcommand{\earr}{\end{array}}
\newcommand{\bi}{\begin{itemize}}
\newcommand{\ei}{\end{itemize}}

\newcommand{\sinc}{\ensuremath{\mathrm{sinc}}}

\usepackage[utf8]{inputenc}
\usepackage[T1]{fontenc} 
\usepackage{lmodern} 
\usepackage{microtype} 
\usepackage{graphicx}
\usepackage{dcolumn}
\usepackage{enumitem,kantlipsum,amsthm,amssymb,color}
\usepackage{amsfonts}
\usepackage{amsmath}
\usepackage{mathtools}
\usepackage{dsfont}
\usepackage{siunitx}

\usepackage{subcaption}
\usepackage{ragged2e}
\DeclareCaptionJustification{justified}{\justifying}
\usepackage{bbm}
\usepackage{bm}
\usepackage{synttree}
\usepackage{float}
\usepackage{bbold}
\usepackage{braket}
\usepackage[table]{xcolor}
\usepackage{gensymb}
\usepackage{tcolorbox}

\usepackage{booktabs}
\usepackage{upgreek, eurosym}
\usepackage[colorlinks=true]{hyperref}
\usepackage[noabbrev]{cleveref}

\newcommand{\fillcol}{\extracolsep{\fill}}

\definecolor{gray}{rgb}{0.5,0.5,0.5}

\graphicspath{{images/}}

\begin{document}


\bibliographystyle{apsrev}


\title{Experimental cheat-sensitive quantum weak coin flipping}

\author{Simon Neves}
\affiliation{Sorbonne Université, CNRS, LIP6, 4 Place Jussieu, Paris F-75005, France}

\author{Verena Yacoub}
\affiliation{Sorbonne Université, CNRS, LIP6, 4 Place Jussieu, Paris F-75005, France}

\author{Ulysse Chabaud}
\affiliation{Institute for Quantum Information and Matter, California Institute of Technology, 1200 E California Blvd, Pasadena, CA 91125, USA}

\author{Mathieu Bozzio}
\affiliation{University of Vienna, Faculty of Physics, Vienna Center for Quantum Science and Technology (VCQ), 1090 Vienna, Austria}

\author{Iordanis Kerenidis}
\affiliation{Université de Paris, CNRS, IRIF, 8 Place Aurélie Nemours, Paris 75013, France}

\author{Eleni Diamanti}
\affiliation{Sorbonne Université, CNRS, LIP6, 4 Place Jussieu, Paris F-75005, France}

%


%


\begin{abstract}

As in modern communication networks, the security of quantum networks will rely on complex cryptographic tasks that are based on a handful of fundamental primitives. Weak coin flipping (WCF) is a significant such primitive which allows two mistrustful parties to agree on a random bit while they favor opposite outcomes. Remarkably, perfect information-theoretic security can be achieved in principle for quantum WCF. Here, we overcome conceptual and practical issues that have prevented the experimental demonstration of this primitive to date, and demonstrate how quantum resources can provide cheat sensitivity, whereby each party can detect a cheating opponent, and an honest party is never sanctioned. Such a property is not known to be classically achievable with information-theoretic security. Our experiment implements a refined, loss-tolerant version of a recently proposed theoretical protocol and exploits heralded single photons generated by spontaneous parametric down conversion, a carefully optimized linear optical interferometer including beam splitters with variable reflectivities and a fast optical switch for the verification step. High values of our protocol benchmarks are maintained for attenuation corresponding to several kilometers of telecom optical fiber.    

\end{abstract}


\maketitle


\noindent Communication network users need to operate or interact with parties, servers, nodes and transmission channels that they do not necessarily trust to handle sensitive data, sign digitally, or perform online banking, delegated computing, and electronic voting, among many other tasks. To guarantee the security of such networking tasks against malicious entities, it is necessary to rely on a collection of building blocks, called cryptographic primitives, which can be combined with one another to guarantee overall security~\cite{BS:dcc16}. Coin flipping is a fundamental primitive that comes in two versions. In strong coin flipping (SCF), two parties remotely agree on a random bit such that none of the parties can bias the outcome with probability higher than $1/2+\epsilon$, where $\epsilon$ is the protocol bias~\cite{Blum:Sigact83}. It is essential for multiparty computation~\cite{GMW:STOC87}, online gaming and more general randomized consensus protocols involving leader election~\cite{AAK:DC18}. In weak coin flipping (WCF), on the other hand, there is a winner and a loser, in the sense that both parties have a preferred, opposite outcome.

In classical communication networks, there exist no secure SCF and WCF protocols without computational assumptions or trusting a third party~\cite{Blum:Sigact83,C:STOC86,A:STOC04,BBB:PRA09}. Although accepting a non-zero abort probability allows for information-theoretically secure classical schemes to exist \cite{HW:TC11}, such schemes cannot detect malicious behaviours deviating from the original protocol. On the other hand, cheat-sensitive coin flipping become possible when using quantum properties. Quantum SCF protocols have in fact been shown to display a fundamental lower bound on their bias~\cite{K:PC03}, but quantum WCF may achieve biases arbitrarily close to zero~\cite{ARV:arx19,ACG:SIAM16}. Interestingly, quantum WCF can also be used for the construction of optimal quantum SCF and quantum bit commitment schemes~\cite{IEEE:CK09,IEEE:CK11}. 

While quantum SCF protocols have been experimentally demonstrated~\cite{MVU:PRL05,BBB:NC11,PJL:NC14}, the implementation of quantum WCF has remained elusive so far, because of the absence of protocols bringing together the use of practical states and measurements with tolerance to losses. Recently, a linear optical implementation, exploiting photon-number encoding, was proposed in~\cite{BCK:PRA20}, but the quantum advantage it can provide in terms of bias is extremely sensitive to losses: a dishonest party may always declare an abort when they are not satisfied with the outcome of the coin flip. Furthermore, an explicit optical implementation of quantum WCF with arbitrarily small bias is yet to be discovered. 

Here, we provide the first, to the best of our knowledge, experimental demonstration of quantum WCF. Our demonstration relies on the generation of heralded single photons by spontaneous parametric down conversion (SPDC), which are effectively entangled with the vacuum on a beam splitter of variable reflectivity. The outcome of the coin flip is then provided by the detection or absence of a photon.
Our protocol is a refined version of the theoretical protocol from~\cite{BCK:PRA20}, which provides a new desirable property in the presence of losses that relates to cheat sensitivity rather than bias: by dropping the condition from~\cite{BCK:PRA20} that both parties have equal probability of winning when cheating, our protocol allows them to detect whether their opponent is cheating during a verification step, and does not sanction an honest party. There are no known classical protocols that achieve such cheat sensitivity~\cite{HK:PRL04,SR:PRL02}.
In order to emphasize the robustness of our protocol to losses, we show that it remains secure over an attenuation that corresponds to several kilometers of telecom optical fiber.


\bigskip

\large
\noindent\textbf{Results}
\normalsize

\noindent\textbf{Protocol.} We first introduce our protocol for quantum weak coin flipping using a single photon, building on the protocol proposed in~\cite{BCK:PRA20}. Our protocol accounts for potential losses and the detection of a cheating party (see Protocol Box). It ends with five mutually incompatible outcomes: Alice wins or is sanctioned, Bob wins or is sanctioned, or the protocol aborts. The protocol uses three beam splitters, whose reflectivities $x$, $y$, and $z$ are chosen in order to satisfy two conditions on these events. Firstly, the \textit{fairness} condition, which states that Alice and Bob have equal winning probabilities when both of them are honest, i.e.\ $\mathbb{P}_h(\textrm{A.\ wins})=\mathbb{P}_h(\textrm{B.\ wins})$, or
\begin{equation}\label{eq:Fair}
    \mathbb{P}_h\bigl[(b,v_1,v_2)=(0,1,0)\bigr]=
    \mathbb{P}_h\bigl[(b,a)=(1,0)\bigr].
\end{equation}
Secondly, the \textit{correctness} condition, which states that an honest party should never be sanctioned for cheating, i.e.\ $\mathbb{P}_h(\textrm{A.\ sanctioned})=\mathbb{P}_h(\textrm{B.\ sanctioned})=0$, or
\begin{equation}\label{eq:correctness}
    \mathbb{P}_h\bigl[(b,v_2) = (0,1)\bigr] = \mathbb{P}_h\bigl[(b,a)=(1,1)\bigr] = 0.
\end{equation}
Note that contrary to the previous protocol~\cite{BCK:PRA20}, we drop the \textit{balancing} condition, which states that Alice and Bob should have equal probabilities of winning when using an optimal cheating strategy. In fact, satisfying this condition implies that the probability of sanctioning an honest Alice is non-negligible, so that the correctness condition is not verified anymore. This impacts the cheat sensitivity, as one cannot trust the verification step if it sanctions honest parties (see Supp.~Mat.~\ref{theoPredictions} for details on the protocol and the chosen conditions).\\

\begin{figure*}[htbp]
	\begin{center}
		\includegraphics[width=170mm]{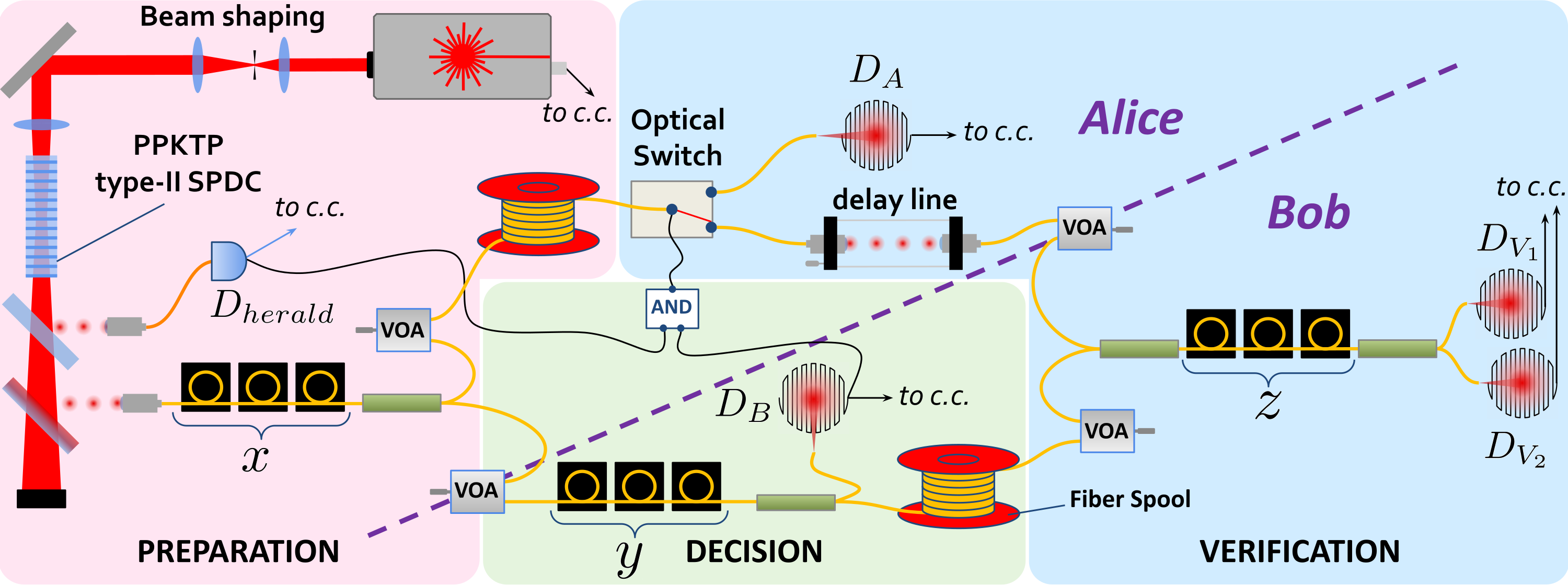}
		\caption{\textbf{Experimental setup for cheat-sensitive quantum weak coin flipping.} A ppKTP crystal ($\SI{30}{mm}$-long, $\SI{46.2}{\micro\meter}$ poling period) is pumped by a $\SI{770}{nm}$ pulsed laser ($\approx\SI{2}{ps}$ long pulses, $\SI{76}{\mega\hertz}$ rate). Twin photons at telecom wavelengths are generated via type-II SPDC, separated from the pump by a dichroic mirror (DM), and from each other by a PBS. The signal photon is used to perform the protocol as shown in the scheme in the Protocol Box, which is the reference for defining the reflectivities $x$, $y$ and $z$. These are tuned with polarization controllers, placed before PBSs. At the end of each of the possible paths, the signal photon is detected with high-efficiency SNSPDs ($D_A, D_B, D_{V_1}, D_{V_2}$). A coincidence counter (c.c.) conditions each detection on the idler photon, detected with an InGaAs avalanche photodiode (APD) $D_{herald}$, and on the emission of a pump pulse, detected via an internal photodiode, measured in a $\SI{500}{ps}$ coincidence window. The signal from Bob's detector, conditioned on the heralding signal via a logic AND gate, triggers a fast optical switch on Alice's side. While these signals are being processed, the photon is delayed by $\SI{300}{m}$-long optical fiber spools on each party's side. A delay line allows for fine-tuning of the wave-packets timing on the last PBS. Communication distance $L$ between Alice and Bob is simulated by VOAs of transmission $e^{-0.02L}$, which are shown on the dashed line marking visually the separation between the two parties. Two more VOAs are included in the setup to simulate losses due to photon storage corresponding to this distance.}
		\label{fig:setup}
	\end{center}
\end{figure*}

\noindent\textbf{Experimental setup.} The experimental setup used for the implementation of our protocol is shown in Fig.~\ref{fig:setup}. Alice generates heralded single photons via type-II SPDC in a periodically-poled potassium titanyl phosphate (ppKTP) crystal. The protocol is implemented using fibered components at telecom wavelength. As polarization is a degree of freedom not used for encoding, Alice entangles it with the spatial modes, using polarizing beam splitters (PBS). In this way, the beam splitters (BS) reflectivities $x$, $y$, and $z$, can be effectively tuned by rotating the single-photon polarization before each PBS, using polarization controllers. We use a fast optical switch in order to select the party who performs the verification step, depending on the outcome $b$. During this operation, the photon is delayed using optical fiber spools; Alice's source together with Bob's verification setup then form a $>\SI{300}{m}$-long fibered Mach-Zehnder interferometer. In order to mitigate the resulting interference noise, we carefully insulated the spools and achieved an interference visibility of $v \gtrsim 96\%$ (see Methods for details).
Under these conditions, the thermally-induced fluctuations are slow enough such that we can easily post-select the protocol runs in which there was no phase difference between the two arms of the interferometer. This post-selection does not threaten the protocol security, as the parties could monitor the interference before performing the coin flip, and agree on starting the protocol only when the phase difference is null. Single photons are detected with threshold superconducting nanowire single-photon detectors (SNSPDs) in order to maximize the detection efficiency. Finally, to simulate communication distance between Alice and Bob, and the corresponding losses induced by the photon storage that is necessary in this case, we use variable optical attenuators (VOAs).

\newcounter{saveenum}
\begin{tcolorbox}[title= Protocol: cheat-sensitive quantum weak coin flipping with a single photon.,title filled]
\begin{center}
\includegraphics[width=70mm]{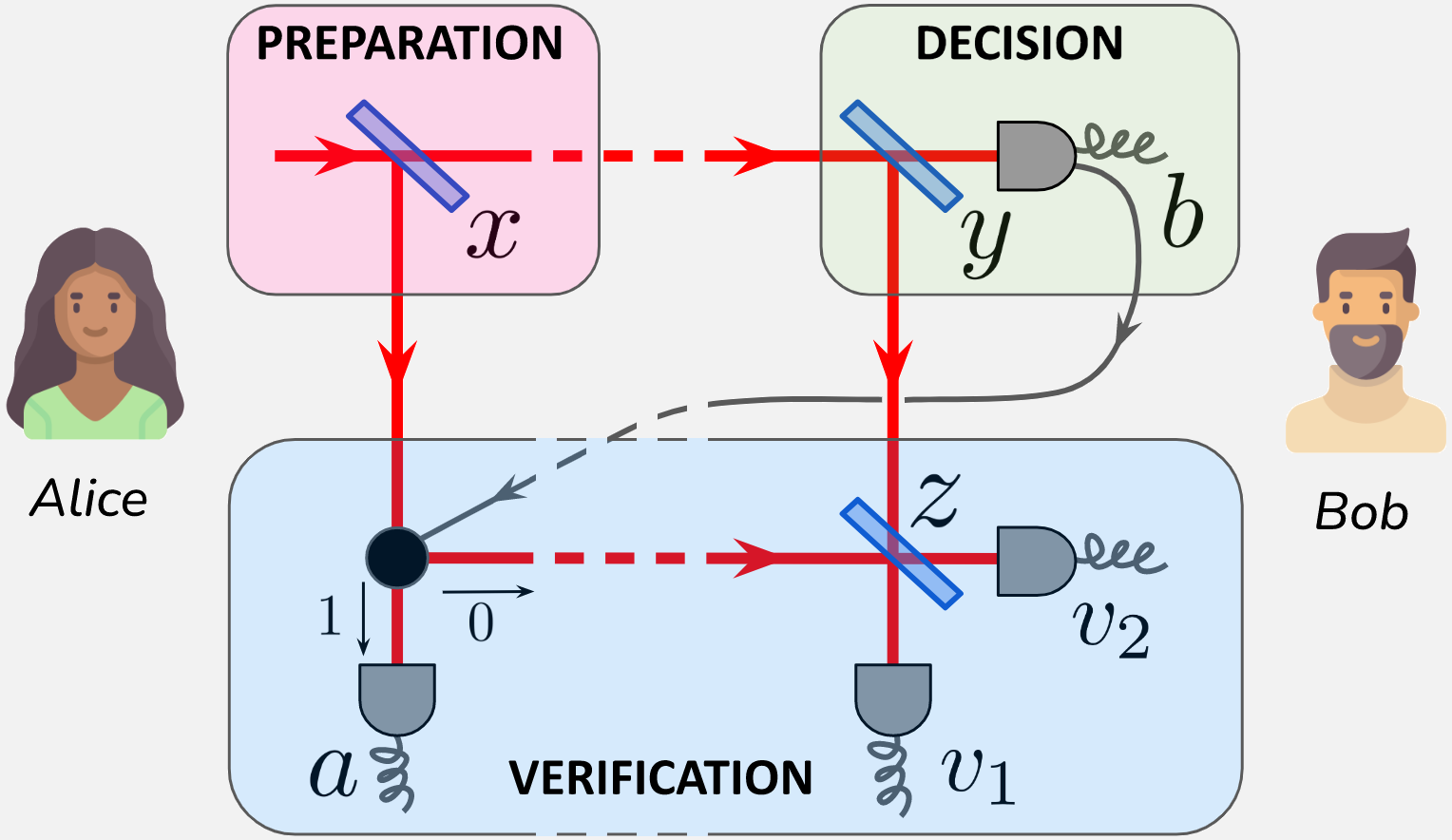}
\end{center}
\begin{enumerate}[wide]
\item\textit{Preparation.} Alice sends a single photon on a beam splitter of reflectivity $x$, keeps the reflected mode, and sends the other to Bob.
\item\textit{Decision.} Bob sends the state he receives on a beam splitter of reflectivity $y$, measures the transmitted mode with a single-photon detector $D_B$, and broadcasts the outcome $b\in\{0,1\}$.
\item\textit{Verification.} If $b = 0$, Alice sends her reflected mode to Bob, who mixes it with his own reflected mode on a beam splitter of reflectivity $z$, and measures the two outputs with single-photon detectors $D_{V_1}$ and $D_{V_2}$. He distinguishes three cases depending on the outcome $(v_1, v_2)$:
    \begin{itemize}
        \item $v_2 = 1$: Alice is sanctioned for cheating,
        \item $(v_1, v_2) = (1, 0)$: Alice wins,
        \item $(v_1,v_2) = (0, 0)$: the protocol aborts.
    \end{itemize}
    
If $b = 1$, Bob discards his state. Alice measures her state with a single-photon detector $D_A$. She discerns two cases depending on the outcome $a$:
    \begin{itemize}
        \item $a=0$: Bob wins,
        \item $a=1$: Bob is sanctioned for cheating.
    \end{itemize}
\end{enumerate}
\vspace{-0.2cm}
\vspace{-0.1cm}
\end{tcolorbox}

\begin{table}[htbp]
\begin{tabular}{|c|c|@\fillcol c @\fillcol|@\fillcol c @\fillcol|@\fillcol c @\fillcol|@\fillcol c @\fillcol|c|}
 \hline
  Notation & Path & \hspace{0.1cm}$x$\hspace{0.1cm} &\hspace{0.0cm} $y$ \hspace{0.1cm}& \hspace{0.1cm}$z$\hspace{0.1cm} &\hspace{0.0cm} $s$\hspace{0.1cm} & Efficiency\\
  \hline
  $\eta_A^s$ & $x \rightarrow \textrm{switch} \rightarrow D_A$& 1 &\cellcolor[rgb]{0.5,0.5,0.5}&\cellcolor{gray}& 1& $0.315\pm 0.008$\\
  \hline 
  $\eta_B^y$& $x \rightarrow y \rightarrow D_B$ & 0 & 0 &\cellcolor{gray}&\cellcolor{gray}&$0.303\pm 0.008$\\
  \hline
  $\eta_A^{V_1}$& $x \rightarrow \textrm{switch} \rightarrow z \rightarrow D_{V_1}$ & 1 &\cellcolor{gray}& 1 & 0&$0.231\pm 0.008$\\
  \hline
  $\eta_A^{V_2}$& $x \rightarrow \textrm{switch} \rightarrow z \rightarrow D_{V_2}$& 1 &\cellcolor{gray}& 0 & 0&$0.219\pm 0.008$\\
  \hline
  $\eta_B^{V_1}$& $x \rightarrow y \rightarrow z \rightarrow D_{V_1}$& 0 & 1 & 0 &\cellcolor{gray}&$0.184\pm 0.008$\\
  \hline
  $\eta_B^{V_2}$& $x \rightarrow y \rightarrow z \rightarrow D_{V_2}$ & 0 & 1 & 1 &\cellcolor{gray}&$0.175\pm 0.008$\\
  \hline
\end{tabular}
\caption{\textbf{List of notations and measured values for the efficiencies corresponding to the different paths involved in the experiment.} The paths are described by the PBSs (labelled by the corresponding reflectivities) and/or the switch they go through, as well as the detector at the end of the path. We also list the values of $x$, $y$, $z$, and the state of the switch $s$, required to measure these efficiencies. Values are given for VOAs set at $\SI{0}{\decibel}$.}\label{tab:Losses}
\end{table}

Because of their central role in the analysis of the protocol, we wish to distinguish the BS reflectivities from the losses induced by the rest of the components in the setup. For that purpose we define different transmission (or heralding) efficiencies, measured when the reflectivities and the state of the switch are set to trivial values $x,y,z,s\in\{0,1\}$. These values reflect the losses in every possible path in the experiment, which are induced for instance by fiber spools, VOAs, fiber coupling and mating, or detectors. We detail the notations for the efficiencies corresponding to each path and their measured values in Table~\ref{tab:Losses}. Each path is defined by the detector it ends in and the arm it goes through (Alice's or Bob's).\\

\noindent\textbf{Results with honest parties.} We are first interested in the protocol when both Alice and Bob are honest. In our experiments, because of dark counts, double-pair emission, or imperfect interference visibility, Alice and Bob can still be sanctioned even though they are honest and the setup is optimized. In general, we cannot tune the reflectivities perfectly, so Alice and Bob may have slightly different winning probabilities. This means our implementation cannot satisfy perfectly the fairness and correctness conditions. Therefore, we define the fairness $\mathcal{F}$ and correctness $\mathcal{C}$ in order to quantify the closeness to these two conditions as follows:
\begin{align}
    &\mathcal{F} = 1- \biggl\vert\dfrac{ \mathbb{P}_h(\textrm{A.\ wins}) - \mathbb{P}_h(\textrm{B.\ wins}) }{\mathbb{P}_h(\textrm{A.\ wins}) + \mathbb{P}_h(\textrm{B.\ wins})}\biggr\vert, \label{eq:FairnessDef}\\
    &\mathcal{C} = 1 -\dfrac{\mathbb{P}_h(\textrm{A.\ sanctioned}) + \mathbb{P}_h(\textrm{B.\ sanctioned})}{\mathbb{P}_h(\textrm{A.\ wins}) + \mathbb{P}_h(\textrm{B.\ wins})}. \label{eq:AccuracyDef}
\end{align}
Both quantities are equal to $1$ when the corresponding conditions are perfectly fulfilled, and $\mathcal{C}, \mathcal{F} < 1$ otherwise. In our implementation, the probability of emitting a pair in a pump pulse is $p\simeq 0.015$, so double-pair emissions are highly unlikely. We condition any detection event on the detection of both a pump pulse and a heralding photon, which effectively minimizes the already low dark count rates in SNSPDs. In this way, we can omit the double-pair emissions and dark counts as a first approximation, such that only the interference visibility $v$ limits $\mathcal{C}$ and $\mathcal{F}$. Under these assumptions, we show that the correctness and fairness conditions are optimally approached by setting the following reflectivities:
\begin{align}
    &x_h= \biggl[1+\dfrac{\eta_A^{V_1}}{\eta_B^{V_1}}+\dfrac{\eta_A^{V_1}}{\eta_B^y}(1+v)\biggr]^{-1},\label{eq:xtheo}\\[0.2cm]
    &y_h= \biggl[1+\dfrac{\eta_B^{V_1}}{\eta_B^y}(1+v)\biggr]^{-1},\label{eq:ytheo}\\[0.2cm]
    &z_h= \dfrac{1}{2}.\label{eq:ztheo}
\end{align}
The reader can refer to Supp.~Mat.~\ref{theoPredictions} for the detailed proof and \ref{reflectivities} for the values used in our implementation. Then, as long as the parties are honest, we obtain the following probabilities for significant events:
\begin{align}
    &\mathbb{P}_h\bigl(\textrm{A.\ wins}\bigr) = \mathbb{P}_h\bigl(\textrm{B.\ wins}\bigr) = x_h\eta_A^{V_1}(1+v),\\
    &\mathbb{P}_h\bigl(\textrm{B.\ sanctioned}\bigr) = 0,\\
    &\mathbb{P}_h\bigl(\textrm{A.\ sanctioned}\bigr) = x_h\eta_A^{V_2}(1-v).\label{eq:ProbaHonestAliceSanctioned}
\end{align}
Note here the importance of maximizing the interference visibility $v$ so that Alice is not sanctioned while being honest. The above expressions also provide a systematic way to optimize the reflectivities for honest parties, which does not require their direct measurement (see Methods for details).

\begin{figure}[htbp]
	\begin{center}
		\includegraphics[width=85mm]{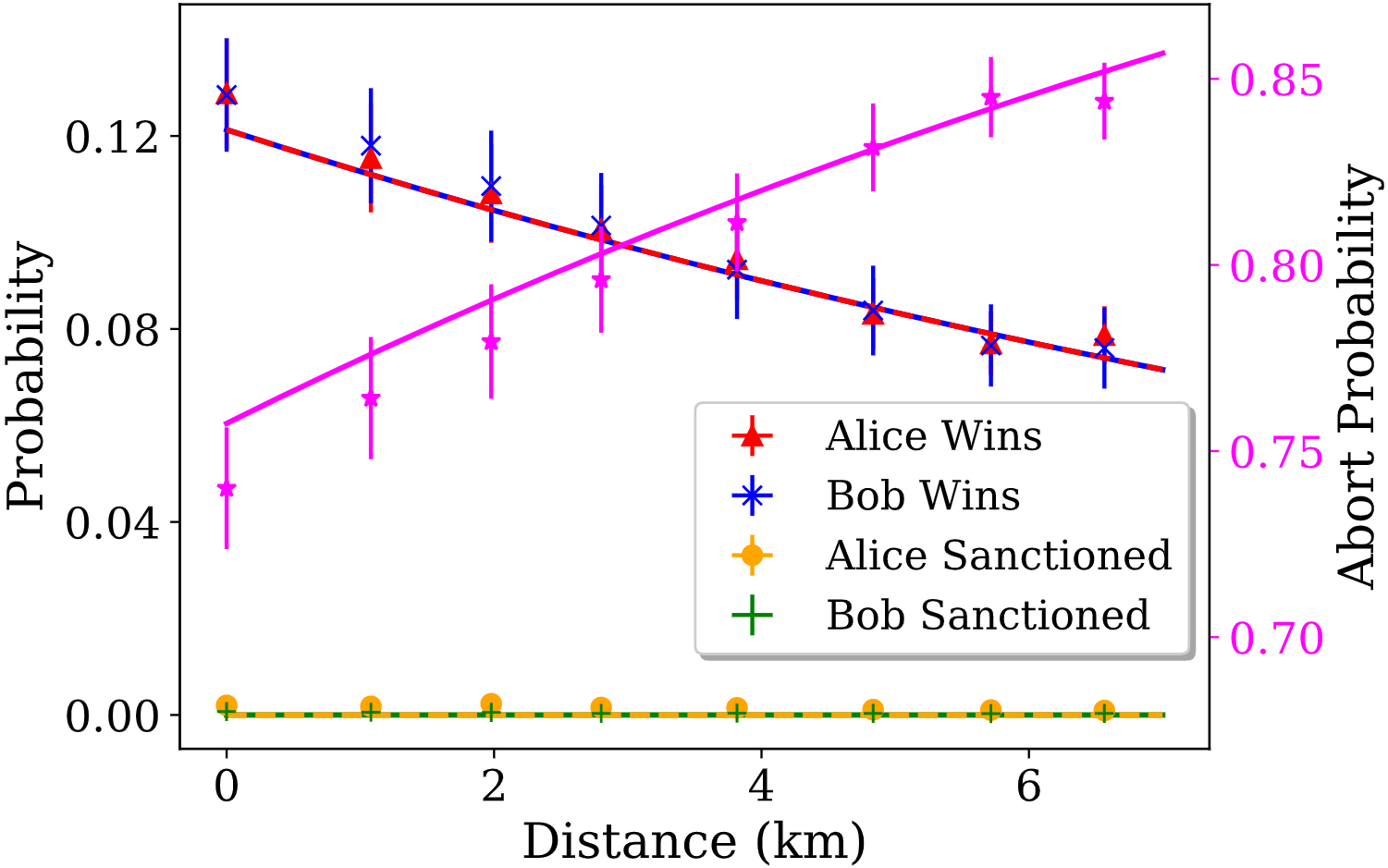}
		\caption{\textbf{Probability of each outcome of the protocol, measured for different communication distances between Alice and Bob.} The abort probability is shown on the right axis, in magenta. The lines represent the theoretical evolution of probabilities, calculated via Eqs.~(\ref{eq:xtheo})~to~(\ref{eq:ProbaHonestAliceSanctioned}), with efficiencies given in Table~\ref{tab:Losses}. The error bars are mainly due to error propagation on these efficiencies.}\label{fig:probaHonests}
	\end{center}
\end{figure}

We perform the protocol for different communication distances between Alice and Bob. These are simulated by setting each of the VOAs to a transmission $\eta = e^{-0.02 L}$ with $L$ the distance in kilometers, introducing additional losses to each arm of the setup. We optimize the fairness and correctness at each distance by tuning the reflectivities. We continuously run the protocol and record all detection events regardless of the phase difference between the two arms of the interferometer. Detection of both a heralding photon and a pump pulse triggers a protocol run. The average protocol rate is $\simeq\SI{51}{\kilo\hertz}$. As if Bob was monitoring the phase difference, we post-select the runs for which the phase spontaneously goes to zero thanks to slow temperature fluctuations, such that the rate in $D_{V_2}$ (which essentially corresponds to the probability of honest Alice being sanctioned) is minimized. In this way, we measure at least $\SI{1.5e5}{}$ valid iterations of the protocol for a $15$-minute run, making the Poisson noise negligible. In Fig.~\ref{fig:probaHonests} we give the probabilities of the different events for several distances. 

\begin{figure}[htbp]
	\begin{center}
		\includegraphics[width=80mm]{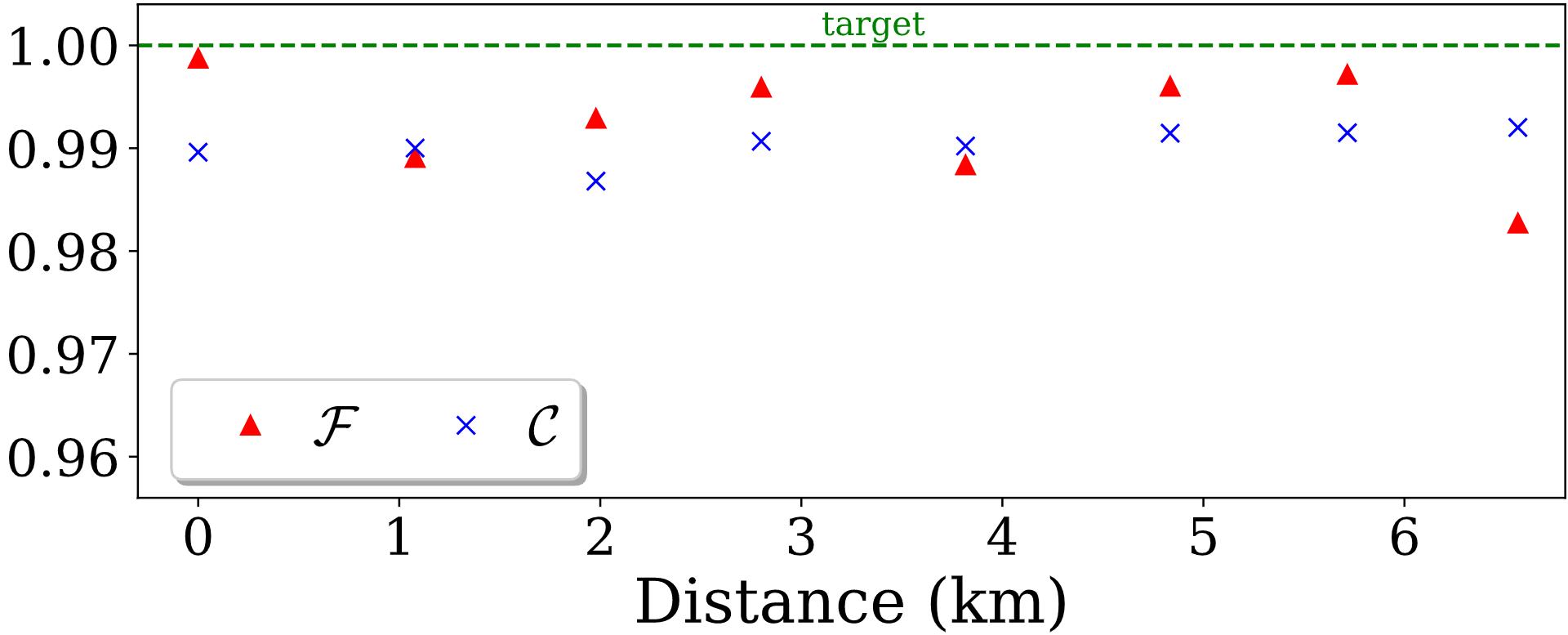}
		\caption{\textbf{Correctness $\mathcal{C}$ and fairness $\mathcal{F}$} measured in our experimental implementation of the protocol with honest parties, for different communication distances. The dashed line gives the target value for an ideal protocol.}\label{fig:FairnessCorrectness}
	\end{center}
\end{figure}

We notice that the abort probability takes relatively high values, even when we trivially set the communication distance to $L=\SI{0}{\kilo\meter}$. This has to do with important losses, particularly in mating sleeves connecting the numerous optical fiber components, the delay line, or in crystalline components such as the PBSs or the optical switch. Significant improvements could be made, using integrated optics for instance. Other critical features are the single-photon coupling and SNSPD efficiencies. Both of these aspects are being actively studied and could see significant improvement in the near future. We also notice that the winning probabilities of Alice and Bob are indeed very close and the probability of an honest party to be sanctioned is minimized. 

To further illustrate the performance of our protocol, we show the fairness $\mathcal{F}$ and correctness $\mathcal{C}$ in Fig.~\ref{fig:FairnessCorrectness}. Thanks to the appropriate tuning of reflectivities $x$, $y$, and $z$, as well as low dark count rates and high visibility, we were able to keep both of these quantities very close to 1, thus approaching the ideal conditions.\\


\noindent\textbf{Cheat sensitivity, results with dishonest parties.} Now we highlight the cheat sensitivity of our protocol, by implementing attacks by dishonest parties. We consider one party to be dishonest, the other one being honest. Bob's optimal cheating strategy is quite straightforward, and consists in claiming $b=1$ regardless of the actual measurement in detector $D_B$~\cite{BCK:PRA20}. As Alice is honest she sets the reflectivity $x=x_h$ given in Eq.~(\ref{eq:xtheo}). When Bob claims $b=1$ then Alice's switch directs her mode in detector $D_A$ so that she can verify whether Bob is being honest. She then detects a photon with probability:
\begin{equation}\label{eq:probaDisBobSanctioned}
    \mathbb{P}(a=1\vert \textrm{B.\ cheats}) = x_h\eta_A^s,
\end{equation}
in which case Bob is sanctioned for cheating. Otherwise, Bob wins with probability:
\begin{equation}\label{eq:probaDisBobwins}
\begin{aligned}
    \mathbb{P}(a=0\vert \textrm{B.\ cheats}) &= 1-\mathbb{P}(a=1\vert \textrm{B.\ cheats})\\&=1-x_h\eta_A^s.
\end{aligned}
\end{equation}
In this way, Alice's conditional verification, enabled in our setup by the fast optical switch, allows for a first kind of cheat sensitivity. 

\begin{figure}[htbp]
	\begin{center}
		\includegraphics[width=80mm]{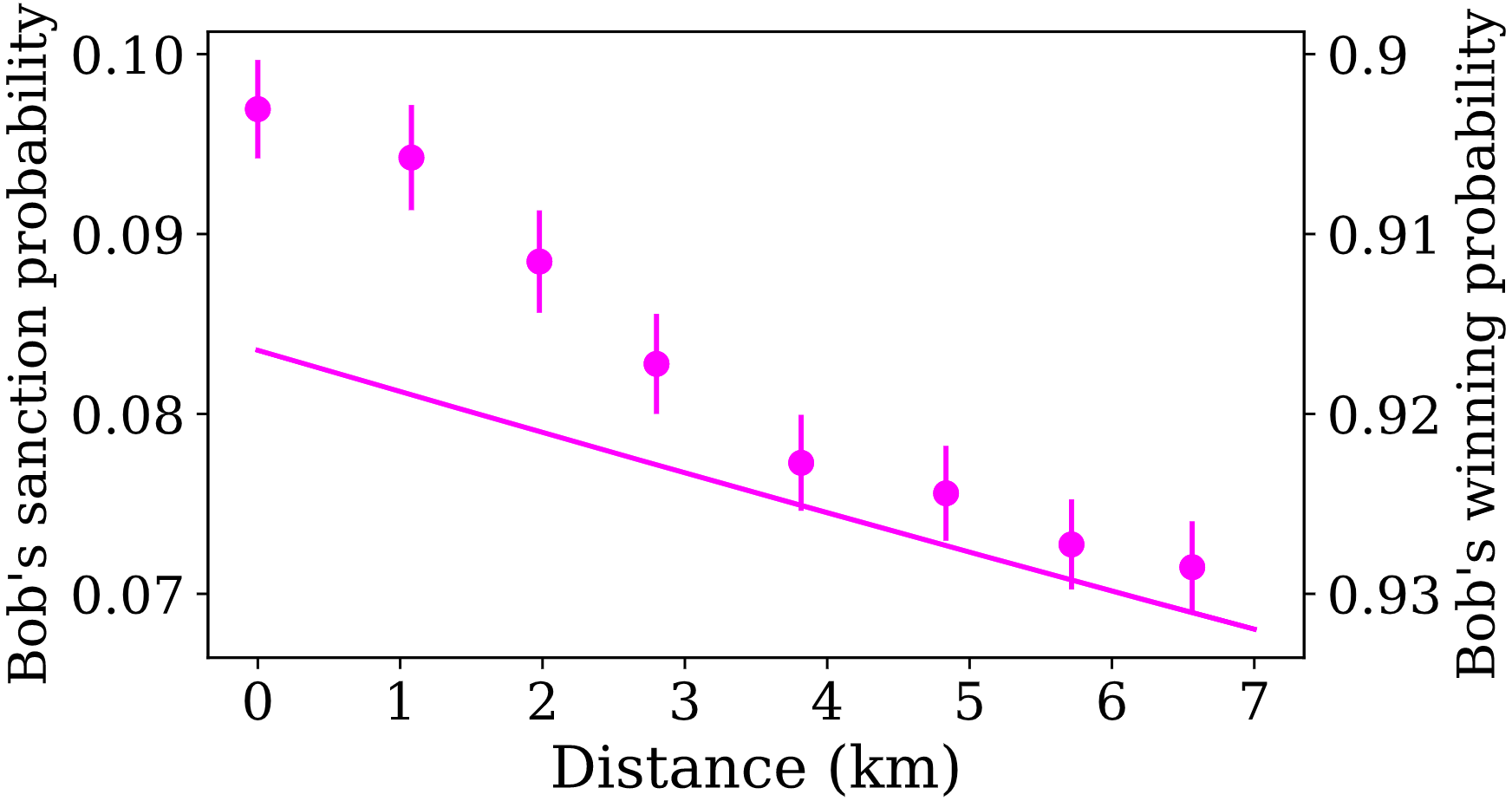}
		\caption{\textbf{Probabilities of Bob winning or being sanctioned, when he is performing an optimal attack, measured for different communication distances between Alice and Bob.} Only one set of points is shown for the two axes, as these two events are complementary. The line is plotted from Eqs.~(\ref{eq:probaDisBobSanctioned}) and~(\ref{eq:probaDisBobwins}), with $\eta_A^s$ given in Table~\ref{tab:Losses}. The error bars are mainly due to error propagation on this efficiency. The observed deviation from the theory is linked to systematic errors when setting the reflectivities, which is discussed in Supp.~Mat.~\ref{reflectivities}.}
		\label{fig:BobCheats}
	\end{center}
\end{figure}

In order to demonstrate this aspect, we implement Bob's optimal strategy by systematically forcing the switch to send the photon to $D_A$. We measure the probability of sanctioning Bob for each of the communication distances simulated in the honest case. As displayed in Fig.~\ref{fig:BobCheats}, we show experimentally that the probability of sanctioning Bob decreases as communication-induced losses increase, therefore limiting Alice's cheat sensitivity. This gives a substantial advantage to Bob when Alice's arm is particularly lossy. Note that when Bob implements that strategy, only two events are possible, namely Bob winning or Bob being sanctioned; Alice can never win except if the sanction is precisely giving Alice the win (see discussion below).

\begin{figure*}[htbp]
\includegraphics[height=55mm]{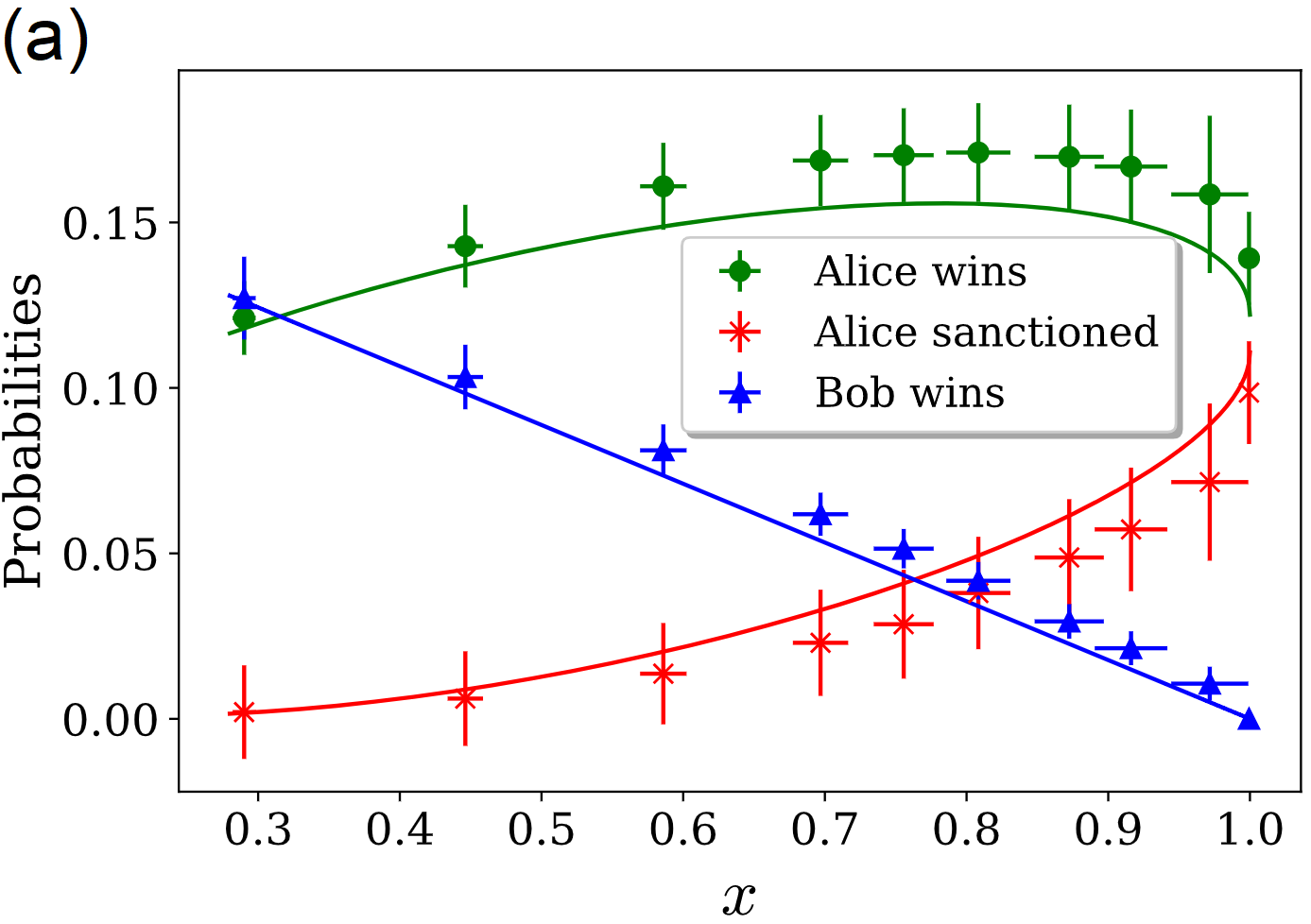}
\hspace{0.5cm}\includegraphics[height=55mm]{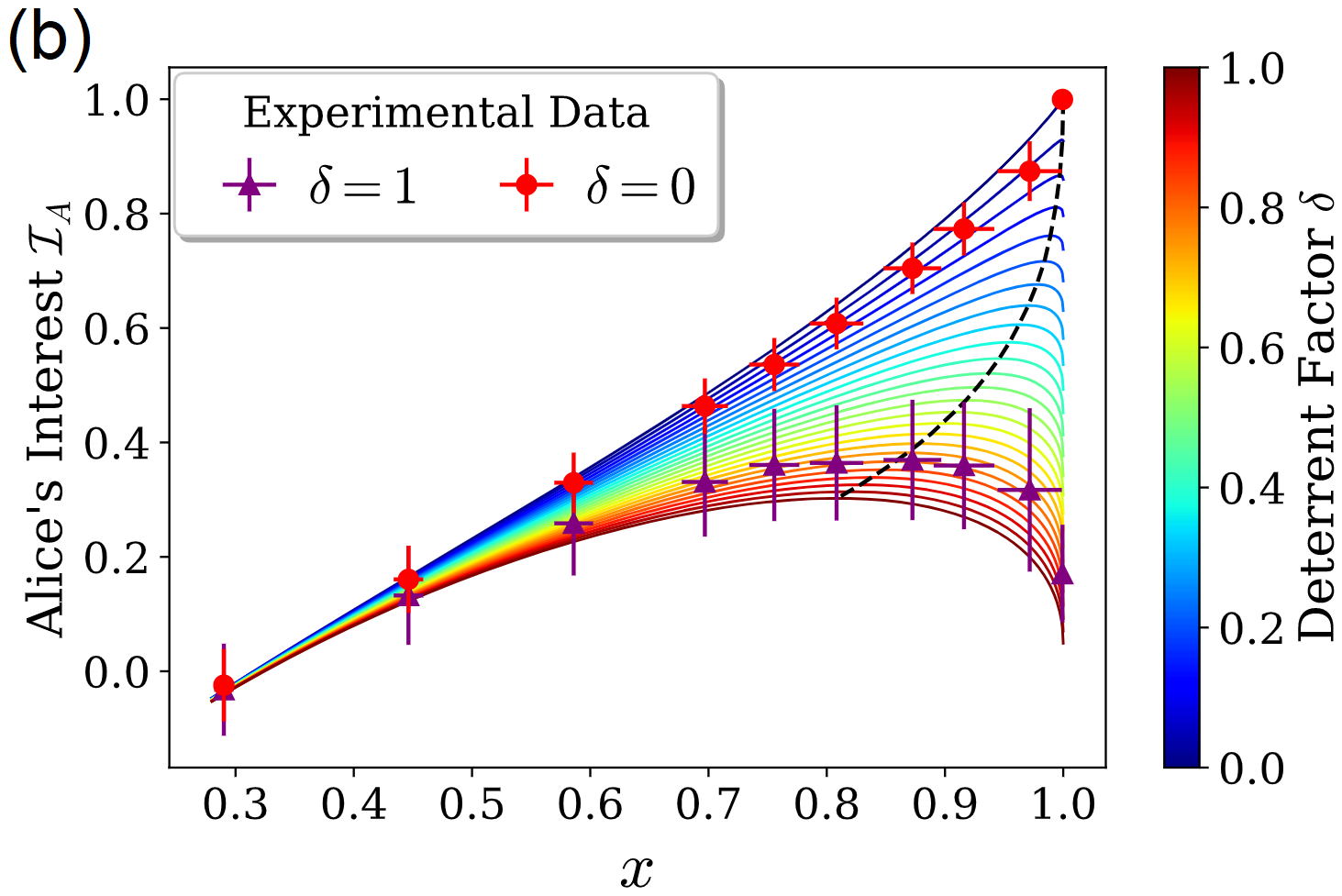}
\caption{\textbf{Results for a dishonest Alice who sets different values of $x$ than the honest value.} The lines show theoretical predictions, calculated from Eqs.~(\ref{eq;AliceCheatsProbasMain})-(\ref{eq:Interest}), with efficiencies given in Table~\ref{tab:Losses}. \textbf{(a)} Probabilities of different outcomes. \textbf{(b)} Alice's cheating interest for different deterrent factors $\delta$. The dashed black line indicates the points of maximum interest.}
		\label{fig:AliceMalicious}
\end{figure*}

On the other hand, when Bob is honest and Alice is dishonest, her optimal cheating strategy is less straightforward, as shown in~\cite{BCK:PRA20}. It relies on the preparation of specific states that are hard to produce with current technology. Additionally, without strong assumptions, complex methods are required in order to find such a strategy, which to this day has not been achieved yet. Still, we can perform suboptimal strategies by simply tuning the value of $x$, so that Alice sends the photon to her side with higher probability: intuitively, without taking the verification setup into account, we can naively expect Alice's winning probability to increase as she increases the reflectivity $x$. We experimentally perform the protocol for different values of $x$, all of them higher than the honest value~(\ref{eq:xtheo}). In that case, the expected event probabilities are given by the following formula (see Supp.~Mat.~\ref{theoPredictions} for the detailed proof): 
\begin{align}\label{eq;AliceCheatsProbasMain}
&\begin{aligned}
    \mathbb{P}(\textrm{A.\ wins})= \dfrac{1}{2}\biggl(x\eta_A^{V_1} + (1-x)y_h\eta_B^{V_1}  &\\&\hspace{-2.3cm} + 2v\sqrt{x(1-x)y_h\eta_A^{V_1}\eta_B^{V_1}}\biggr),
\end{aligned}\\
&\begin{aligned}
    \mathbb{P}(\textrm{A.\ sanctioned})= \dfrac{1}{2}\biggl(x\eta_A^{V_2} + (1-x)y_h\eta_B^{V_2}&\\&\hspace{-3.2cm}  - 2v\sqrt{x(1-x)y_h\eta_A^{V_2}\eta_B^{V_2}}\biggr),
\end{aligned}\\
&\mathbb{P}(\textrm{B.\ wins})= (1-x)(1-y_h)\eta_B^y. 
\end{align}
In Fig.~\ref{fig:AliceMalicious}(a), we show the probabilities of significant events. Contrary to our naive conjecture, we see that thanks to Bob's verification, and thus cheat sensitivity, Alice does not have a clear interest in forcing $x=1$, as her winning probability peaks around $x\simeq 0.78$.

Alice's interest in cheating actually depends on how deterrent the sanction is. We define a factor $\delta \geq 0$, which quantifies that deterrability, or alternatively how harmful the sanction is for a cheating party. From this parameter we can derive an empirical function that quantifies Alice's interest in cheating:
\begin{equation}\label{eq:Interest}
     \mathcal{I}_A(\delta)\!=\!\dfrac{\mathbb{P}(\textrm{A.\ wins})-\mathbb{P}(\textrm{B.\ wins})-\delta\mathbb{P}(\textrm{A.\ sanctioned})}{\mathbb{P}(\textrm{A.\ wins})+\mathbb{P}(\textrm{B.\ wins})+\delta\mathbb{P}(\textrm{A.\ sanctioned})}.
\end{equation}
This function is built such that it can be linked to the fairness~(\ref{eq:FairnessDef}) when taking the appropriate sanction. Indeed, if for $\delta \in [0,1]$ we sanction a cheating Alice by giving the win to Bob with probability $\delta$, then the relation $\mathcal{F} = 1-\vert\mathcal{I}_A(\delta)\vert$ holds. In this way, $\delta = 0$ corresponds to a protocol that simply aborts without sanction when Alice is caught, and $\delta = 1$ gives a protocol that always declares Bob the winner when Alice is caught. Ultimately $\mathcal{I}_A(\delta)$ can be interpreted as a sort of expectation value of a cheating Alice, or a comparison between what she can gain by cheating and what she can lose. In Fig.~\ref{fig:AliceMalicious}(b) we plot Alice's cheating interest for different values of $\delta$ and $x$. If no sanction is taken ($\delta=0$), we see that her interest in cheating grows with $x$. Indeed, even if her winning probability decreases for high values of $x$, Bob's then approaches zero, such that Alice wins with absolute certainty as long as the protocol does not abort. On the contrary, as the sanction is tightened and the value of $\delta$ increases, Alice has less interest in cheating for a given value of $x$. Furthermore, the value of $x$ that maximizes $\mathcal{I}_A$ also goes down, showing how strengthening the sanction actually forces Alice to adopt a strategy that leaves a chance for Bob to win.\\

\large
\noindent\textbf{Discussion}
\normalsize

\noindent After refining a previous theoretical proposal for a practical quantum weak coin flipping protocol \cite{BCK:PRA20}, we were able to perform the first implementation of this protocol by generating a heralded single photon, and entangling it effectively with the vacuum. Thanks to the use of low dark counts SNSPDs, tunable beam splitters and a fast optical switch, while keeping a high visibility in our fibered interferometer, we demonstrated a fair and cheat-sensitive protocol. Importantly, this last property allows to detect a cheating party with non-negligible probability.

Note that in order to sanction a dishonest party with high probability, one could systematically sanction the winning party, regardless of their honesty. Thus, in order to display genuine cheat sensitivity, we highlight the primary importance of the correctness condition, which ensures an honest party is never sanctioned for cheating. This forced us to ignore the balancing of the benefit gained by each party when adopting an optimal cheating strategy, which was previously assessed as a necessary condition for a weak coin flipping protocol \cite{BCK:PRA20}. Still, we propose a way of restoring this balance, by using the deterrent factor and interest function introduced in the previous paragraphs.

The balance could indeed arise from choosing different sanctions for Alice and Bob, associated with different deterrent factors $\delta_A$ and $\delta_B$, in order to equalize the corresponding interest functions $\mathcal{I}_A(\delta_A)$ and $\mathcal{I}_B(\delta_B)$. A dishonest party who could dramatically increase their winning probability would therefore take a bigger risk of being harshly sanctioned when cheating. Interestingly enough, one could actually set arbitrarily big deterrent factors $\delta > 1$ in order to account for harsher sanctions. We leave the evaluation of these sanctions, deterrent factors and potential alternative interest functions as an interesting game theory open question.

From an experimental perspective, we remark that the robustness to losses in our implementation was illustrated by simulating communication distance with variable optical attenuators. In a practical implementation of the protocol, it would be necessary to maintain a high visibility for a longer interferometer, which could be achieved with active stabilization techniques used in twin-field quantum key distribution implementations for instance~\cite{CMD:NC22,CZJ:PRL20}.\\

\large
\noindent\textbf{Methods}
\normalsize

\noindent\textbf{Source and detection.} Our single-photon signal was heralded by its idler twin, in a pair generated via type-II SPDC in a ppKTP crystal (Raicol). We maximized the heralding efficiency $\eta_s = R_{si} / R_i$, with $R_i$ the idler photon detection rate and $R_{si}$ the pair detection rate. For that purpose, the pump focus and pair collection modes were tuned carefully when coupling to single-mode fibers, and losses on the signal-photon path were minimized. In particular, we used $>\SI{85}{\percent}$-efficiency SNSPDs (ID281 from ID Quantique) to detect that photon. Losses on the idler photon were not limiting, so we detected it with a $\SI{25}{\percent}$-efficiency InGaAs APD (ID230 from ID Quantique). In this way, without adding the rest of the components, we measured a maximum heralding efficiency $\eta_s = 63\%$. All detection events were recorded by a time tagger  (Time Tagger Ultra from Swabian Instruments), and dated with picosecond precision. Two detection events were considered simultaneous when measured in a $\SI{500}{ps}$ coincidence window. We also use the pump laser as a clock, in order to filter out most of the dark counts from the APD, which occur at a $\SI{1}{\kilo\hertz}$-rate. In this way, protocol runs were triggered at a rate of $\SI{51}{\kilo\hertz}$, with $\SI{40}{\hertz}$ of runs mistakenly triggered by dark counts.

\medskip

\noindent\textbf{Error management.} Various factors can generate undesired detection events in our protocol. This is true in particular for sanction outcomes, triggered by a detection in $D_A$ or $D_{V_2}$ which should never occur when a party is honest. Most of these outcomes arise from Bob's verification procedure, which relies on a Mach--Zehnder interferometer. If this interference is of poor visibility, then $D_{V_2}$ can be triggered even if Alice is being honest, and her winning probability is also substantially lowered. Considering the length of this interferometer ($>\SI{300}{m}$), the visibility is limited by two main factors, namely the coherence length and phase fluctuations. The coherence length of photons is $\simeq \SI{2.4}{\milli\meter}$, which is small enough to start losing coherence after a few hours of experiment runs. This is mostly caused by length variations in the interferometer arms due to thermal fluctuations ($\simeq \SI{2.4}{\milli\meter/\celsius}$ for a $\SI{300}{m}$ arm). We therefore regularly tune the length of one arm of the interferometer, using a free-space micro-metric delay line. 
Phase fluctuations can be separated into two regimes. Slow phase fluctuations, of typical frequency $\lesssim\SI{1}{\hertz}$, are again caused by thermal variations. We can easily measure them, and then either correct them or simply post-select the desired phase differences. Fast phase fluctuations, however, are caused by noise spanning the audible spectrum from $\SI{20}{\hertz}$ to $\SI{2}{\kilo\hertz}$. This noise is amplified by the $\SI{300}{m}$ fiber spools, which act as sort of microphone. These fluctuations are hard to resolve with our single-photon rate of a few $\SI{10}{\kilo\hertz}$, such that the interference pattern is averaged on that noise, and we witness an interference visibility of approximately $v\simeq 80\%$. In order to characterize that noise, we measure the interference pattern with a continuous diode laser and a fast photodiode. Without any sound insulation, the noise in the interference fluctuation spans the audible spectrum with a power spectral density of approximately $\simeq\SI{7e-3}{\volt^2/\hertz}$. In order to mitigate this noise, we wrap the fiber spools into several layers of sound-absorbing floating parquet underlay. We then achieve an interference visibility of $v\simeq 96\%$.

\medskip

\noindent\textbf{Reflectivity setting.} When the parties are honest, Bob first sets $z=1/2$ by blocking Alice's signal, and equalizing the detection rates in $D_{V_1}$ and $D_{V_2}$. This later ensures an optimized interference, and therefore the correctness condition. Then he can tune $y$ such that the detection rate in $D_B$ equals twice the total rate in $D_{V_1}$ and $D_{V_2}$, which should ensure the fairness condition. Alice then tunes $x$ in order to optimize the interference visibility, which should complete the setting of reflectivities. If $v$ is significantly lower than $1$, Alice and Bob might have to perform some mild adjustments on $x$ and $y$ in order to maximize the fairness and correctness. After performing a protocol with reflectivities $x,y,z$ we can evaluate them by measuring some specific probabilities (see Supp.~Mat. for more details).

\medskip

\noindent\textbf{Optical switching.} During the decision step of the protocol, Bob's detection determines which party is winning, and which one is performing the verification. This decision is effectively taken into account by Alice via her optical switch (Nanospeed from Agiltron). In this way, if Bob does not claim victory, the switch is in state "0" in order to send Alice's state to Bob, who performs the verification. If Bob claims victory, the switch goes to state "1" such that Alice keeps her state and performs the verification. In practice, we send the electronic signal from Bob's detector, together with the heralding signal, to a fast programmable logic AND gate, integrated in a time controller (ID900 from ID Quantique). This AND gate filters out potential detection events outside of the protocol, which might saturate the optical switch. The gate's output signal is then sent to the optical switch, which executes the decision (see Supp.~Mat. for more details). 

\medskip

\large
\noindent\textbf{Acknowledgments}\\
\normalsize
We acknowledge financial support from the European Research Council project QUSCO (E.D.) and the European Commission project QUANGO. U.C.\ acknowledges funding provided by the Institute for Quantum Information and Matter, an NSF Physics Frontiers Center (NSF Grant PHY-1733907).

\bigskip
\bigskip
\bigskip

\large
\noindent\textbf{Author contributions}\\
\normalsize
S.N.,\ M.B.\ and E.D.\ designed and S.N.\ developed the experimental setup. S.N.\ and V.Y.\ performed the protocol implementation and processed the data. S.N.,\ V.Y.,\ U.C.\ and M.B.\ performed the protocol analysis. All authors discussed the analysis of the data, and contributed to writing or proofreading the manuscript. I.K.\ and E.D.\ supervised the project.

\medskip

\large
\noindent\textbf{Additional information}\\
\normalsize
The authors declare no competing financial interests.


\newpage
\onecolumngrid

\appendix

\section*{Supplementary Material}

\section{Theoretical predictions}\label{theoPredictions}

\noindent In this section we give some theoretical predictions for the results we observe in our experiments. In the first two subsections, we derive general expressions for event probabilities, for any values of beam splitter reflectivities $x$, $y$, and $z$. In the third subsection we obtain the values of these reflectivities which maximize fairness and correctness, when both parties are honest, as well as the probabilities of the different outcomes. In the fourth subsection, we show such predictions when one of the parties, Alice, is dishonest and performs an attack which we implement in this paper. In general, these predictions differ from those derived in previous work~\cite{BCK:PRA20}, as we drop the \textit{balancing} condition for the correctness, and we adopt a different parametrization. We give some development on that matter in the last subsection

\subsection{Photon propagation in the interferometer}

\begin{figure}[htbp]
	\begin{center}
		\includegraphics[width=120mm]{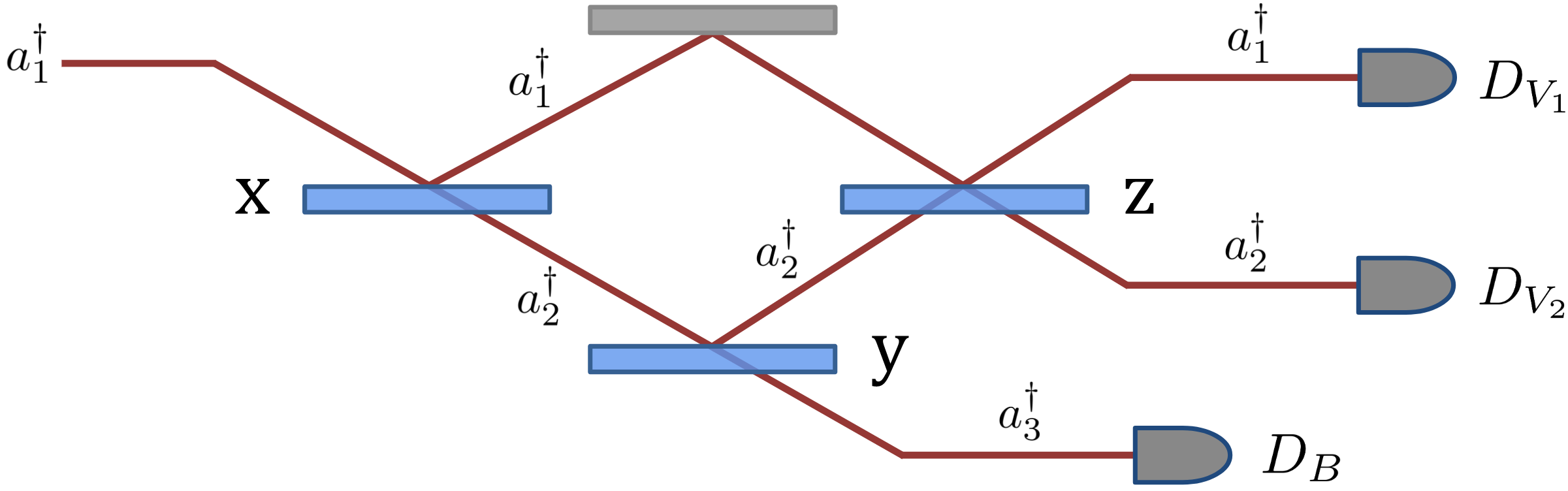}
		\caption{Sketch of the interferometer with most relevant notations.}
		\label{fig:SchematicSetup}
	\end{center}
\end{figure}

\noindent We first describe the propagation of the photon in the interferometer (see Fig.~\ref{fig:SchematicSetup}), for any values of $x$, $y$, $z$, and deduce the probabilities of the different events. To simplify our proofs, we neglect dark counts and double-pair emissions. Experimental details in the following paragraphs support the legitimacy of this approximation. In this scenario, when Alice detects a photon in detector $D_{herald}$, then exactly one photon is generated, corresponding to the action of the creation operator $a_1^\dagger$. Some first losses occur when coupling the photon to single-mode fibers, such that the operator transforms as:
\begin{equation}
    a_1^\dagger \longrightarrow \sqrt{\eta_c} \: a_1^\dagger,
\end{equation}
with $\eta_c$ being the induced transmission. Then Alice sends the photon to a BS of reflectivity $x$:
\begin{equation}
   \sqrt{\eta_c} \: a_1^\dagger \longrightarrow \sqrt{x\eta_c} \: a_1^\dagger + \sqrt{(1-x)\eta_c}\: a_2^\dagger,
\end{equation}
where 1 (resp.\ 2) stands for the reflected (resp.\ transmitted) mode. Alice keeps mode 1 and Bob gets mode 2. On each side, the photon undergoes losses due to fiber transmission and connectors, storage, and diverse other components. We note $\eta_{A1}$ (resp.\ $\eta_{B1}$) the transmission on Alice's (resp.\ Bob's) side. Some phases are also induced by the propagation, and we note $\Phi_{A1}$ (resp.\ $\Phi_{B1}$) the phase introduced on Alice's (resp.\ Bob's) side. In this way, we get the following transformation:  
\begin{equation}
   \sqrt{x\eta_c}\: a_1^\dagger + \sqrt{(1-x)\eta_c}\: a_2^\dagger \longrightarrow \sqrt{x\eta_c\eta_{A1}} \: e^{i\Phi_{A1}} a_1^\dagger + \sqrt{(1-x)\eta_c\eta_{B1}}\: e^{i\Phi_{B1}} a_2^\dagger.
\end{equation}
Bob sends the photon to a BS of reflectivity $y$:
\begin{equation}
\begin{aligned}
   \sqrt{x\eta_c\eta_{A1}} \: e^{i\Phi_{A1}} a_1^\dagger + \sqrt{(1-x)\eta_c\eta_{B1}}\: e^{i\Phi_{B1}} a_2^\dagger &\\&\hspace{-4cm}\longrightarrow \sqrt{x\eta_c\eta_{A1}} \: e^{i\Phi_{A1}} a_1^\dagger + \sqrt{(1-x)y\eta_c\eta_{B1}}\: e^{i\Phi_{B1}} a_2^\dagger + \sqrt{(1-x)(1-y)\eta_c\eta_{B1}}\: e^{i\Phi_{B1}} a_3^\dagger.
\end{aligned}
\end{equation}
Bob sends the third mode to the detector $D_B$, inducing another loss. We note $\eta_y$ the transmission, including the detector efficiency, and we have $\eta_B^y = \eta_c \eta_{B1} \eta_y$ (here we omit the dephasing as no interference will occur on this mode). The second mode undergoes some loss and dephasing, and we note $\eta_{B2}$ and $\Phi_{B2}$ the transmission and dephasing. There we note $\eta_B = \eta_c \eta_{B1} \eta_{B2}$ the total loss on Bob's arm of the interferometer, and $\Phi_B = \Phi_{B1} + \Phi_{B2}$ the total dephasing. On Alice's side, the path depends on the detection of the third mode that triggers the optical switch. In absence of dark counts and when Bob is honest, a detection on the third mode means no detection will occur on Alice's verification detector, such that Bob is not sanctioned and wins the coin flip. In other words, Alice trusts Bob's measurement on the third mode, such that we can omit her verification detector and the optical switch. In that case she simply sends the first mode to Bob to proceed to verification of the state. That mode undergoes some loss and dephasing, and we note $\eta_{A2}$ and $\Phi_{A2}$ the transmission and dephasing. There we note $\eta_A = \eta_c \eta_{A1} \eta_{A2}$ the total loss on Alice's arm of the interferometer, and $\Phi_A = \Phi_{A1} + \Phi_{A2}$ the total dephasing. The total transformation becomes:
\begin{equation}
\begin{aligned}
   \sqrt{x\eta_c\eta_{A1}} \: e^{i\Phi_{A1}} a_1^\dagger + \sqrt{(1-x)y\eta_c\eta_{B1}}\: e^{i\Phi_{B1}} a_2^\dagger + \sqrt{(1-x)(1-y)\eta_c\eta_{B1}}\: e^{i\Phi_{B1}} a_3^\dagger &\\&\hspace{-7cm}\longrightarrow \sqrt{x\eta_A} \: e^{i\Phi_A} a_1^\dagger + \sqrt{(1-x)y\eta_B}\: e^{i\Phi_B} a_2^\dagger + \sqrt{(1-x)(1-y)\eta_B^y}\: e^{i\Phi_B} a_3^\dagger.
\end{aligned}
\end{equation}
After receving the first mode, Bob makes it interfere with the second mode on a BS of reflectivity $z$, such that we get:
\begin{equation}
\begin{aligned}
    \sqrt{x\eta_A} \: e^{i\Phi_A} a_1^\dagger + \sqrt{(1-x)y\eta_B}\: e^{i\Phi_B} a_2^\dagger + \sqrt{(1-x)(1-y)\eta_B^y}\: e^{i\Phi_B} a_3^\dagger  &\\&\hspace{-7cm}\longrightarrow  (\sqrt{xz\eta_A} \: e^{i\Phi_A} +\sqrt{(1-x)y(1-z)\eta_B}\: e^{i\Phi_B} )\: a_1^\dagger \\&\hspace{-6cm}- (\sqrt{x(1-z)\eta_A} \: e^{i\Phi_A}-\sqrt{(1-x)yz\eta_B}\: e^{i\Phi_B})\: a_2^\dagger + \sqrt{(1-x)(1-y)\eta_B^y}\: e^{i\Phi_B} a_3^\dagger.
\end{aligned}
\end{equation}
Bob sends the first and second modes to detectors $D_{V_1}$ and $D_{V_2}$, with efficiencies $\eta_{V_1}$ and $\eta_{V_2}$, and we note $\eta_A^{V_1} = \eta_A \: \eta_{V_1}$, $\eta_A^{V_2} = \eta_A \: \eta_{V_2}$, $\eta_B^{V_1} = \eta_B \: \eta_{V_1}$, and $\eta_B^{V_2} = \eta_B \: \eta_{V_2}$. Up to an irrelevant global phase $e^{i\Phi_A}$, we get:
\begin{equation}
\begin{aligned}
 (\sqrt{xz\eta_A} \: e^{i\Phi_A} +\sqrt{(1-x)y(1-z)\eta_B}\: e^{i\Phi_B} )\: a_1^\dagger &\\&\hspace{-6cm}- (\sqrt{x(1-z)\eta_A} \: e^{i\Phi_A}-\sqrt{(1-x)yz\eta_B}\: e^{i\Phi_B})\: a_2^\dagger + \sqrt{(1-x)(1-y)\eta_B^y}\: e^{i\Phi_B} a_3^\dagger 
     \\&\hspace{-4cm}\longrightarrow  \bigl(\sqrt{xz\eta_A^{V_1}} \:  +\sqrt{(1-x)y(1-z)\eta_B^{V_1}}\: e^{i\Delta\Phi} \bigr)\: a_1^\dagger \\&\hspace{-3cm}- \bigl(\sqrt{x(1-z)\eta_A^{V_2}} \:-\sqrt{(1-x)yz\eta_B^{V_2}}\: e^{i\Delta\Phi}\bigr)\: a_2^\dagger + \sqrt{(1-x)(1-y)\eta_B^y}\: e^{i\Delta\Phi} a_3^\dagger, 
\end{aligned}
\end{equation}
where $\Delta\Phi=\Phi_B-\Phi_A$ is the phase difference. We deduce the detection probabilities in each detector:
\begin{align}
    &P_{V_1}=\mathbb{P}_h\bigl((b,v_1,v_2)=(0,1,0)\bigr)= xz\eta_A^{V_1} + (1-x)y(1-z)\eta_B^{V_1} + 2\cos(\Delta \Phi)\sqrt{x(1-x)yz(1-z)\eta_A^{V_1}\eta_B^{V_1}},\\
    &P_{V_2}=\mathbb{P}_h\bigl((b,v_2)=(0,1)\bigr)= x(1-z)\eta_A^{V_2} + (1-x)yz\eta_B^{V_2} - 2\cos(\Delta \Phi)\sqrt{x(1-x)yz(1-z)\eta_A^{V_2}\eta_B^{V_2}},\label{eq:Pv2}\\
    &P_{D_B}=\mathbb{P}_h\bigl((b,a)=(1,0)\bigr)= (1-x)(1-y)\eta_B^y.\label{eq:PvBproof}
\end{align}

\subsection{Phase fluctuations}

\noindent In our experiment, the phase difference $\Delta\Phi$ evolves through time, because of thermal fluctuations and diverse vibrations or noise. Slow phase drifts, typically caused by thermal fluctuations, are generally resolved when counting photons, provided the photon rate is high enough. Fast phase fluctuations however, typically caused by noise, are hard to resolve by counting photons, due to low rates and detector recovery time. Hence, the probabilities $P_{V_1}$ and $P_{V_2}$ are averaged over the typical temporal resolution $\tau$ of our detectors. We distinguish two types of behaviour in the phase difference $\Delta\Phi(t) = \Delta\Phi_f(t) + \Delta\Phi_s(t)$, with $\Delta\Phi_f(t)$ corresponding to fast fluctuations of typical period $\tau_f \ll \tau$, and $\Delta\Phi_s(t)$ corresponding to slow fluctuations of typical period $\tau_s \gg \tau$. For fast fluctuations, the average value $\langle \cos\Delta\Phi_f\rangle_\tau$ is approximately constant. For slow fluctuations, the value of $\Delta\Phi_s(t)$ is approximately constant over a time lapse of $\tau$. In this way, we get:
\begin{equation}
    \begin{aligned}
    \langle \cos\Delta\Phi\rangle_\tau(t) &= \langle \cos(\Delta\Phi_f+\Delta\Phi_s)\rangle_\tau(t)\\
                                    &= \langle \cos\Delta\Phi_f\cos\Delta\Phi_s\rangle_\tau(t)-\langle \sin\Delta\Phi_f\sin\Delta\Phi_s\rangle_\tau(t_0)\\
                                    &= \langle \cos\Delta\Phi_f\rangle_\tau\cos\Delta\Phi_s(t)-\langle \sin\Delta\Phi_f\rangle_\tau\sin\Delta\Phi_s(t)\\
                                    &= v\: \bigl( {C} \cdot \cos\Delta\Phi_s(t)- {S} \cdot \sin\Delta\Phi_s(t)\bigr),
    \end{aligned}
\end{equation}
with $v := \sqrt{\langle \cos\Delta\Phi_f\rangle_\tau^2+\langle \sin\Delta\Phi_f\rangle_\tau^2}$, ${C} :=  \langle\cos\Delta\Phi_f\rangle_\tau/v$, and ${S} := \langle \sin\Delta\Phi_f\rangle_\tau/v$. By definition we have ${C}^2 + {S}^2 = 1$, so there exists a phase $\phi_{\textrm{eff}}$ with ${C} = \cos \phi_{\textrm{eff}}$ and ${S} = \sin \phi_{\textrm{eff}}$. We then get:
\begin{equation}
    \begin{aligned}
    \langle \cos\Delta\Phi\rangle_\tau(t) &= v\: \bigl( \cos \phi_{\textrm{eff}} \cos\Delta\Phi_s(t)- \sin \phi_{\textrm{eff}} \sin\Delta\Phi_s(t)\bigr)\\
    &= v\: \cos(\Delta\Phi_s(t)+ \phi_{\textrm{eff}}).
    \end{aligned}
\end{equation}
Here $\phi_{\textrm{eff}}$ appears as an additional constant dephasing, such that we can include it inside the slow dephasing $\Delta\Phi_s(t_0)$. Effectively, it means taking $\phi_{\textrm{eff}} = 0$, such that ${S} = 0$ and $\langle \sin\Delta\Phi_f\rangle_\tau = 0$. In this way, we have:
\begin{equation}
    \langle \cos\Delta\Phi\rangle_\tau(t) = v\: \cos\Delta\Phi_s(t)\,
\end{equation}
with $v = \vert\langle \cos\Delta\Phi_f\rangle_\tau\vert \in [0,1]$, that we later interpret as the interference visibility. Now we average $P_{V_1}$ and $P_{V_2}$:
\begin{align}
    &\langle P_{V_1}\rangle_\tau(t)= xz\eta_A^{V_1} + (1-x)y(1-z)\eta_B^{V_1} + 2v\cos\bigl(\Delta\Phi_s(t)\bigr)\sqrt{x(1-x)yz(1-z)\eta_A^{V_1}\eta_B^{V_1}},\label{eq:Pv1phase}\\
    &\langle P_{V_2}\rangle_\tau(t)= x(1-z)\eta_A^{V_2} + (1-x)yz\eta_B^{V_2} - 2v\cos\bigl(\Delta\Phi_s(t)\bigr)\sqrt{x(1-x)yz(1-z)\eta_A^{V_2}\eta_B^{V_2}},\label{eq:Pv2phase}
\end{align}
which are the effective expressions of $P_{V_1}$ and $P_{V_2}$ we can use for our estimations in the following. For this reason, we omit the averaging and time dependence in the remainder of the paper.

\subsection{Predictions with honest parties}

\noindent We now consider a protocol where both parties are honest, and we derive the parameters $x$, $y$ and $z$ that maximize the fairness and correctness. The fairness condition imposes:
\begin{equation}
    \mathbb{P}_h\bigl((b,a)=(1,0)\bigr) = \mathbb{P}_h\bigl((b,v_1,v_2) = (0,1,0)\bigr),
\end{equation}
and the correctness condition imposes:
\begin{equation}
    \mathbb{P}_h\bigl((b,a)=(1,1)\bigr) = \mathbb{P}_h\bigl((b,v_2) = (0,1)\bigr) = 0.
\end{equation}
As we neglected dark counts and double-pair emissions, we already have $\mathbb{P}_h\bigl((b,a)=(1,1)\bigr) = 0$. However, we have \textit{a priori} $\mathbb{P}_h\bigl((b,v_2) = (0,1)\bigr) > 0$ for any non-trivial parameters $x,y,z \notin \{0,1\}$ (these cases do not allow to verify the fairness condition). It is therefore impossible in principle to verify the correctness condition. Still, we minimize $\mathbb{P}_h\bigl((b,v_2) = (0,1)\bigr)$ in order to approach the condition. As a reminder, we have:
\begin{equation}
    \mathbb{P}_h\bigl((b,v_2) = (0,1)\bigr) = P_{V_2} = x(1-z)\eta_A^{V_2} + (1-x)yz\eta_B^{V_2} - 2v\cos\bigl(\Delta\Phi_s\bigr)\sqrt{x(1-x)yz(1-z)\eta_A^{V_2}\eta_B^{V_2}}.
\end{equation}
We first notice that minimizing that expression imposes $\Delta\Phi_s = 0$. Now we recall that $\eta_A^{V_2} = \eta_A\eta_{V_2}$ and $\eta_B^{V_2} = \eta_B\eta_{V_2}$, and define $\Pi_A = x\eta_A$ and $\Pi_B = (1-x)y\eta_B$ that we interpret as the probabilities of measuring the photon in Alice's side or Bob's side, before the last tunable BS. We can then rewrite the probability:
\begin{equation}
     P_{V_2} = \eta_{V_2}\cdot \biggl((1-z)\Pi_A + z\Pi_B - 2v\sqrt{z(1-z)\Pi_A\Pi_B}\biggr).
\end{equation}
We can then define a variable $\xi :=\tfrac{\Pi_A}{\Pi_\textrm{tot}}\in [0,1]$ with $\Pi_\textrm{tot} = \Pi_A+\Pi_B$, such that:
\begin{equation}
     P_{V_2} = \eta_{V_2}\Pi_\textrm{tot}\biggl((1-z)\xi + z(1-\xi)- 2v\sqrt{z(1-z)\xi(1-\xi)}\biggr).
\end{equation}
$P_{V_2}$ is minimized for $\partial P_{V_2} / \partial \xi = 0$ and $\partial P_{V_2} / \partial z = 0$. One can easily show that this imposes $\xi = z = 1/2$, such that $\Pi_A = \Pi_B$. This drastically simplifies the expressions of the probabilities:
\begin{align}
    &P_{V_1}= x\eta_A^{V_1}(1+v),\\
    &P_{V_2}= x\eta_A^{V_2}(1-v),\\
    &\hspace{-0.15cm}x\eta_A^{V_1} = (1-x)y\eta_B^{V_1}.\label{eq:proofParam1}
\end{align}
Now we can apply the fairness condition, which in the honest case with no dark counts and no double-pair emission reduces to $P_{V_1} = P_{D_B}$. This gives the following equation on the parameters:
\begin{equation}\label{eq:proofParam2}
    x\eta_A^{V_1}(1+v) = (1-x)(1-y)\eta_B^y.
\end{equation}
Combining Eqs.~(\ref{eq:proofParam1}) and (\ref{eq:proofParam2}), we can derive the expressions of the three parameters $x$, $y$ and $z$ that optimize both fairness and correctness:
\begin{align}
    &x_h= \biggl[1+\dfrac{\eta_A^{V_1}}{\eta_B^{V_1}}+\dfrac{\eta_A^{V_1}}{\eta_B^y}(1+v)\biggr]^{-1},\\[0.2cm]
    &y_h= \biggl[1+\dfrac{\eta_B^{V_1}}{\eta_B^y}(1+v)\biggr]^{-1},\\[0.2cm]
    &z_h= \dfrac{1}{2}.
\end{align}
Then, the probabilities of the different events are calculated straightforwardly:
\begin{align}
    &\mathbb{P}_h\bigl(\textrm{Alice wins}\bigr) = \mathbb{P}_h\bigl(\textrm{Bob wins}\bigr) = P_{V_1} = P_{D_B} = x_h\eta_A^{V_1}(1+v),\\
    &\mathbb{P}_h\bigl(\textrm{Bob sanctioned}\bigr) = 0,\\
    &\mathbb{P}_h\bigl(\textrm{Alice sanctioned}\bigr) = P_{V_2} = x_h\eta_A^{V_2}(1-v).
\end{align}
This confirms that the correctness condition is not fulfilled in general, but is approached when $v$ gets close to $1$, i.e.\ when the noise is low enough. We detail the experimental procedure for that noise cancellation in a later section. One can also notice that the condition $\Pi_A = \Pi_B$, which later translates to Eq.~(\ref{eq:proofParam1}), gives the expected result that the two arms of the interferometer should have equal power in order to display an optimized interference. This should be kept in mind when experimentally setting up the parameters. Finally, by keeping the same reflectivities, and comparing the values of $P_{V_1}$ and $P_{V_2}$ when $\Delta\Phi_s = 0$ or $\Delta\Phi_s = \pi$, we get:
\begin{equation}
    v = \biggl\vert\dfrac{P_{V_1}(\Delta\Phi_s = 0)-P_{V_1}(\Delta\Phi_s = \pi)}{P_{V_1}(\Delta\Phi_s = 0)+P_{V_1}(\Delta\Phi_s = \pi)}\biggr\vert = \biggl\vert\dfrac{P_{V_2}(\Delta\Phi_s = 0)-P_{V_2}(\Delta\Phi_s = \pi)}{P_{V_2}(\Delta\Phi_s = 0)+P_{V_2}(\Delta\Phi_s = \pi)}\biggr\vert,
\end{equation}
so we can indeed interpret $v$ as the interference visibility, which can be easily evaluated experimentally. Finally, we mention that each path's transmission efficiency can be measured by setting the reflectivities and switch's state to trivial values $x,y,z,s\in\{0,1\}$ given in Table~\ref{tab:Losses} of the main text, in which we also give the experimentally measured values of these efficiencies. From these efficiencies we can compute the above theoretically predicted reflectivities $x_h$, $y_h$ and $z_h$, which maximize the fairness $\mathcal{F}$ and correctness $\mathcal{C}$. The evolution of these values with the communication distance are shown in a later section, in Fig.~\ref{fig:reflectivities}, together with the reflectivities measured in our experiments.

\subsection{Predictions for a dishonest Alice}

\noindent Now we derive results for the case when Alice is dishonest and Bob is honest. In general, Alice might be able to perform more sophisticated strategies, involving more complex quantum states, such as those mentioned in~\cite{BCK:PRA20}. Yet, finding optimal cheating strategies for Alice remains an open question. Here we only consider a naive strategy, by simply setting up a reflectivity $x > x_h$, which \textit{a priori} favors Alice. As Bob is honest, we still keep $y=y_h$ and $z=z_h=1/2$ from Eqs.~(\ref{eq:ytheo}) and (\ref{eq:ztheo}), and Alice's verification setup is not required. In that case we can derive the expressions for the probabilities of the different events:
\begin{align}\label{eq;AliceCheatsProbas}
&\mathbb{P}(\textrm{A.\ wins}) = \langle P_{V_1}\rangle = \dfrac{1}{2}\biggl(x\eta_A^{V_1} + (1-x)y_h\eta_B^{V_1} + 2v\sqrt{x(1-x)y_h\eta_A^{V_1}\eta_B^{V_1}}\biggr),\\
&\mathbb{P}(\textrm{A.\ sanctioned})=\langle P_{V_2}\rangle= \dfrac{1}{2}\biggl(x\eta_A^{V_2} + (1-x)y_h\eta_B^{V_2} - 2v\sqrt{x(1-x)y_h\eta_A^{V_2}\eta_B^{V_2}}\biggr),\\
&\mathbb{P}(\textrm{B.\ wins})= P_{D_B} = (1-x)(1-y_h)\eta_B^y.
\end{align}
These come straightforwardly from Eqs.~(\ref{eq:PvBproof}), (\ref{eq:Pv1phase}) and (\ref{eq:Pv2phase}), by noting that a dishonest Alice would still set $\Delta \Phi_s=0$, which maximizes her winning probability and minimizes her sanction probability. This gives the curves plotted in Fig.~\ref{fig:AliceMalicious}(a) in the main text.

\subsection{On the balancing condition}

\noindent We designed our protocol building on the proposal from~\cite{BCK:PRA20}. However, as mentioned in the main text, here we drop the \textit{balancing} condition, which states that Alice and Bob should have the same probability of winning when using an optimal cheating strategy. Indeed, we show in the following that using this condition is not compatible with cheat sensitivity, as a balanced protocol might sanction honest parties for cheating, so we cannot trust the verification procedure. Let us for instance consider a completely lossless setup with perfect interference. Then it was shown in~\cite{BCK:PRA20} that in order to fulfill both the fairness and balancing condition, the reflectivities should be:
\begin{align}
&x = 1 - \dfrac{1}{\sqrt{2}},\\
&y = \dfrac{1}{\sqrt{2}},\\
&z = 2-\sqrt{2}.
\end{align}
Plugging these values in Eq.~(\ref{eq:Pv2}) we compute the probability of Alice being sanctioned for cheating while being honest, $P_{V_2} = 0.03$, which is non negligible. Instead, we choose the \textit{correctness} condition, imposing this probability $P_{V_2}$ to be minimized, in order to enforce the cheat sensitivity. In this specific ideal case, the correctness and fairness conditions give $x = 1/4$, $y = 1/3$, $z = 1/2$. This gives $P_{V_2}=0$, so we can trust Bob's verification apparatus. Maximizing the correctness also increases the probability of successfully nominating a winner when both parties are honest. Indeed, this probability is only limited by the losses and is therefore equal to 
\begin{equation}
    \mathbb{P}_h(\textrm{Alice wins}) + \mathbb{P}_h(\textrm{Bob wins}) = \dfrac{1}{2} + \dfrac{1}{2} = 1,
\end{equation}
when there is no losses in the setup. In the same situation, the balancing condition only gives a success probability of $0.97$. Hence, the success probability of the balanced protocol is always intrinsically limited, while that of the protocol with optimized correctness will be maximized as technology improves.


\section{Experimental details}

\subsection{Heralded single-photon source}

\noindent The heralded single photons are emitted via type-II spontaneous parametric down-conversion (SPDC) in a ppKTP crystal. The energy conservation and quasi-phase matching imposes the following relations between the pump, signal and idler frequencies and spatial momenta:
\begin{align}
    &\Delta\omega = \omega_p - \omega_s - \omega_i = 0\\
    &\Delta k = k_p - k_s - k_i - \dfrac{2\pi}{\Lambda}=0,
\end{align}
where $\omega_p$, $\omega_s$, and $\omega_i$ are the pump, signal and idler frequencies, $k_p$, $k_s$, and $k_i$ are their spatial momenta and $\Lambda$ is the crystal poling period. The pump laser center wavelength is $\lambda_p=\SI{770}{\nano\meter}$, and the poling period is $\Lambda = \SI{46.2}{\micro\meter}$. With these parameters, at room temperature, using Sellmeier's equations from~\cite{fradkin1999tunable,konig2004extended} for ppKTP's optical indices  $n_p= 1.76$, $n_s = 1.73$, and $n_i=1.82$, the signal and idler photon wavelengths are around $\lambda_s \simeq\SI{1541.5}{\nano\meter}$ and $\lambda_i \simeq \SI{1538.5}{\nano\meter}$. 

\begin{figure}[htbp]
\centering
\includegraphics[height=70mm]{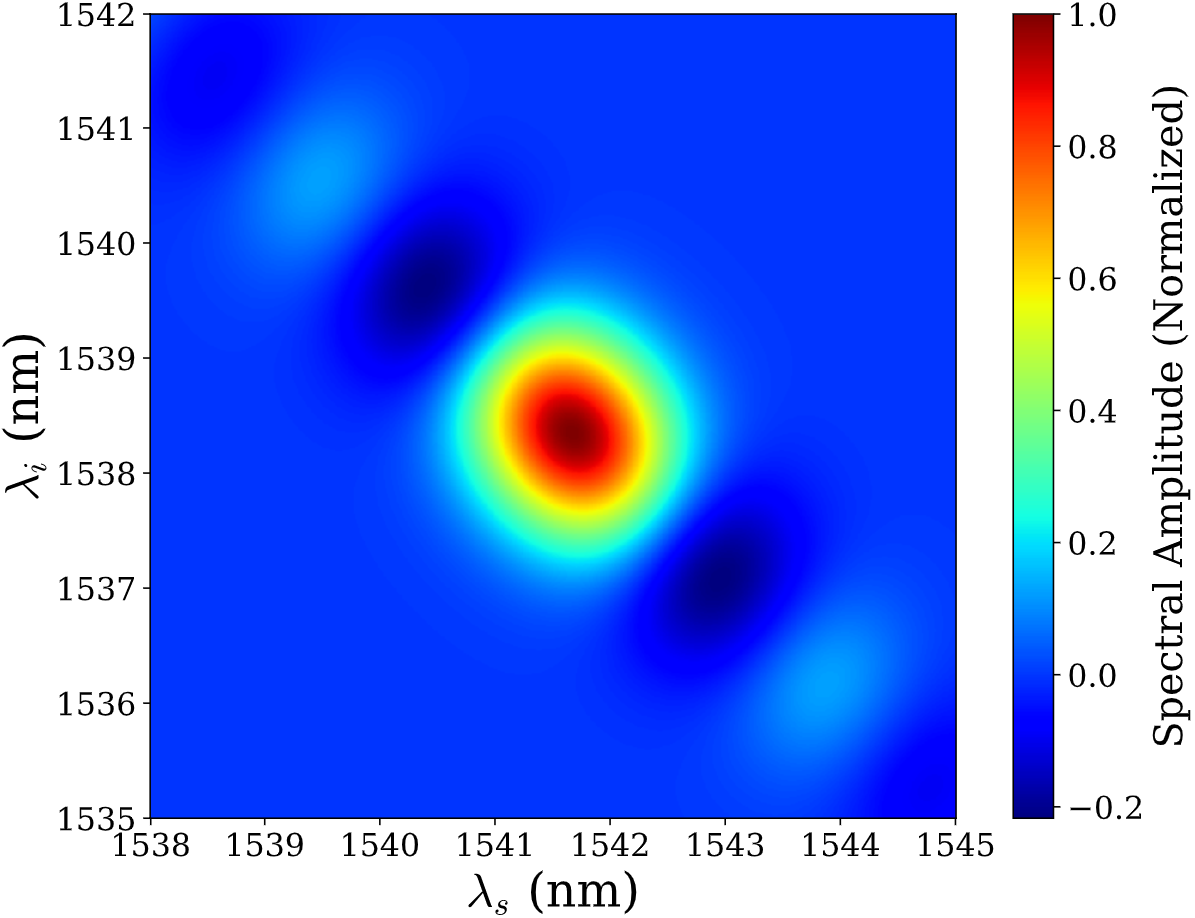}
\caption{Simulated joint spectral amplitude of the photon pairs emitted in our experiment.}
		\label{fig:JSI}
\end{figure}

The pump focus and spectral bandwidth, as well as the crystal length, are of particular relevance to determine or optimize key properties of the photons, such as their coherence length or the heralding efficiency. The pump is focused on the middle of our $L = \SI{30}{\milli\meter}$-long crystal, with a waist of $w_p\simeq\SI{315}{\micro\meter}$. The signal photon's coupling mode has a waist $w_s \simeq \SI{190}{\micro\meter}$, and the idler photon's is $w_i \simeq \SI{218}{\micro\meter}$. In this way, we get the focusing parameters $\xi_p = \tfrac{\pi w_p^2 n_p}{\lambda_p L}\simeq 24$ for the pump, $\xi_s = \tfrac{\pi w_s^2 n_s}{\lambda_s L}\simeq 4.2$ for the signal, and $\xi_i = \tfrac{\pi w_i^2 n_i}{\lambda_i L}\simeq 5.9$ for the idler. We see that the pump beam, as well as photon modes, are close to collimated on the crystal scale. Under these conditions we can consider the spatial state to be uncorrelated from the spectral state~\cite{bennink2010optimal,bruno2014pulsed}. The spectral state of the photons is given by the following expression:
\begin{equation}\label{eq:SpectralState}
    \ket{\psi_{si}} = \dfrac{1}{\mathcal{N}}\iint d\omega_s d\omega_i \gamma(\omega_i,\omega_s) a^\dagger_{\omega_s} a^\dagger_{\omega_i}\ket{0_s,0_i},
\end{equation}
where $\mathcal{N}$ is a normalization factor, and $\gamma(\omega_i,\omega_s)$ is the so-called \textit{joint spectral amplitude} (JSA), which takes the form:
\begin{equation}\label{eq:JSA}
    \gamma(\omega_i,\omega_s) = \alpha(\omega_i+\omega_s)\cdot\phi(\omega_i,\omega_s)
    \vspace{-0.1cm}
\end{equation}
where $\alpha(\omega)=\exp\bigl(-\tfrac{(\omega-\omega_p)^2}{2\sigma_p^2}\bigr)$ is the pump spectrum with $\omega_p$ the central frequency and $\sigma_p$ the bandwidth, $\phi(\omega_s,\omega_i) = \sinc(\Delta k /L)$ is the phase matching amplitude. From the JSA we can extract the Schmidt number $K$ of the pair, and the purity $P$ of each of the photons using $P = 1/K$. Considering the properties of our crystal and our pump Laser's bandwidth $\sigma_p = \SI{0.2}{\nano\meter}$, we expect a purity $P\simeq 0.85$, with a JSA shown in Fig.~\ref{fig:JSI}. Under these conditions, the spectral state is close to pure such that we can evaluate the spectral FWHM $\SI{1}{nm}$ of the single photons as well as their coherence length $\simeq \SI{2.4}{mm}$.

\subsection{Error management}

\noindent Different factors can generate undesired detection events in our protocol, therefore triggering outcomes that would otherwise be of low probability. This is true in particular for sanction outcomes, triggered by a detection in detector $D_A$ or $D_{V_2}$ which should never happen when a party is honest. Thus managing these error sources is of major importance in order to satisfy the correctness condition in particular, but also to minimize undesired outcomes in general. Most of these outcomes arise from Bob's verification procedure, which relies on a Mach--Zehnder interferometer. If this interference is of poor visibility, then $D_{V_2}$ can be triggered even if Alice is being honest, and her winning probability is also substantially lowered. Considering the length of this interferometer ($\geq \SI{300}{m}$), the visibility is limited by two main factors, namely the coherence length and phase fluctuations.

The coherence length of photons is $\simeq \SI{2.4}{\milli\meter}$, which is small enough to start losing coherence after a few hours of experiments. This is mostly caused by length variations in the interferometer arms due to thermal fluctuations ($\simeq \SI{2.4}{\milli\meter/\celsius}$ for a $\SI{300}{m}$ arm). We therefore regularly fine tune the length of one arm of the interferometer, using a free-space micro-metric delay line.

\begin{figure}[htbp]
	\begin{center}
		\includegraphics[width=120mm]{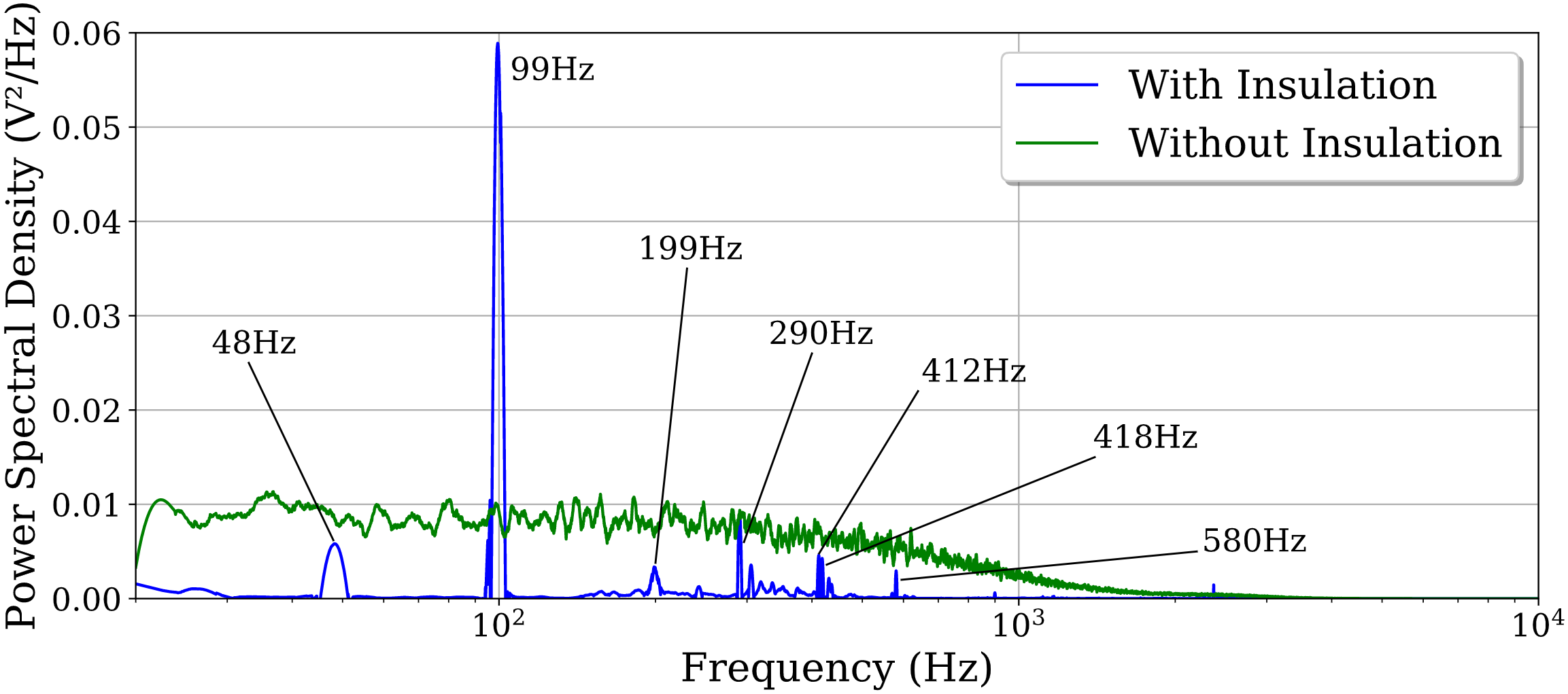}
		\caption{Noise spectrum measured in the interferometer, using a continuous laser and a fast photodiode, with and without sound insulation on the fiber spools. When adding the insulation, the noise is low enough to distinguish peaks coming from the main sources of sound in the lab: $\SI{48}{\hertz}$, $\SI{99}{\hertz}$, and $\SI{199}{\hertz}$ are emitted by the compressor plugged to the detectors cryostat,  $\SI{290}{\hertz}$ and $\SI{580}{\hertz}$ are emitted by the pump in the cold water circuit, which cools down the compressor, and $\SI{412}{\hertz}$ and $\SI{418}{\hertz}$ are emitted by the laser chiller.}
		\label{fig:spectrum}
	\end{center}
\end{figure}

In first approximation, phase fluctuations can be separated into two regimes. Slow phase fluctuations, of typical frequency $\lesssim\SI{1}{\hertz}$, are again caused by thermal variations. We can easily measure them, and then either correct them or simply post-select the desired phase differences. We adopt the latter method in our experiment, which does not threaten the security of the protocol, as parties are allowed to monitor the phase in real time and make the protocol start only when it is set at $\Delta\Phi_s = 0$. Fast phase fluctuations, however, are caused by noise spanning the audible spectrum from $\SI{20}{\hertz}$ to $\SI{2}{\kilo\hertz}$. This noise is amplified by the $\SI{300}{m}$ fiber spools, which act as sort of microphone. These fluctuations are hard to resolve with our single-photon rate of a few $\SI{10}{\kilo\hertz}$, such that the interference pattern is averaged on that noise, and we witness an interference visibility of approximately $v\simeq 80\%$. In order to characterize that noise, we measure the interference pattern with a continuous diode laser and a fast photodiode (see Fig.~\ref{fig:spectrum}). Without any sound insulation, the noise in the interference fluctuation spans the audible spectrum with a power spectral density of approximately $\simeq\SI{7e-3}{\volt^2/\hertz}$. In order to mitigate this effect, we wrap the fiber spools into several layers of sound-absorbing floating parquet underlay. The power spectral density then drops to less than $\SI{1e-3}{\volt^2/\hertz}$ except for some specific frequencies. The total noise power is divided by a factor greater than $\gtrsim 11$. The measured visibility then reaches $v\gtrsim 96\%$. 

Undesired outcomes can also be triggered by double-pair emission inside the crystal, and dark counts in the detectors. The double-pair emission rate is easily minimized by lowering the pump power, such that the probability $p$ of producing a photon pair in a pump pulse is lower than $0.1$. In this way, the probability of generating two pairs in the same pulse is $p^2 \ll p$, such that double-pair emission becomes negligible. In our experiment, we evaluate $p \simeq 0.015$. Dark counts rates are made particularly low by detecting the heralded single photon with SNSPDs with $<\SI{100}{\hertz}$-dark count rate, as well as $<\SI{50}{\pico\second}$-timing jitter electronics. We use the pump internal signal in order to synchronize a $\SI{500}{\pico\second}$ detection gate with each of the detectors signal. Furthermore, all signal-photon detections are conditioned on a heralding photon detection. The probability of detecting a dark count during a protocol run is then $5\cdot10^{-8}$, such that undesired outcomes due to dark counts are negligible. However, we still use an APD, with substantially higher dark count rate than SNSPDs, in order to detect the heralding photon. Such dark counts trigger protocol runs while no photon was emitted. This results in a slight increase of the abort probability, as the other detectors will not click in such a situation. We evaluate the rate of such runs to be $\lesssim \SI{40}{\hertz}$, thanks to the gating applied by the pump laser signal. This way the surplus of abort probability caused by dark count is about $8\cdot 10^{-4}$, which is negligible compared to the typical $>0.7$ abort probability.

\subsection{Optical switch and decision}

\noindent During the decision step of the protocol, Bob's detection determines which party is winning, and which one is performing the verification. In our experiment, this decision is effectively taken into account by Alice via her optical switch. Hence, if Bob does not claim victory, the switch is in state "0" in order to send Alice's state to Bob, who performs the verification. If Bob claims victory, the switch goes to state "1" such that Alice keeps her state and performs the verification. In practice, we send the electronic signal from Bob's detector, together with the heralding signal, to a fast programmable logic AND gate, integrated in a time controller. This AND gate filters out potential detection events outside of the protocol, which might saturate the optical switch. The gate's output signal is then sent to the optical switch, which executes the decision. 

Two timings must be set carefully in order to send the photon in the appropriate direction. First, the two detection electronic signals must be synchronized inside the AND gate in order to perform the logic operation. These timings can be tuned by programming the time controller, and we check that synchronization by measuring the rate of coincidences between the AND gate output, and the detections in the heralding detector and in Bob's detectors. Second, the wave-packet on Alice's side must pass through the switch when the latter is in the appropriate state. As it takes approximately $\simeq \SI{800}{ns}$ to perform the logic gate and the potential shift of the optical switch's state, we use $\SI{300}{m}$-long optical fiber spools, on each party's side, in order to delay the photon for $\simeq\SI{1.5}{\micro\second}$. We can then tune the timing of the AND gate's output electronic signal, again by programming the time controller, so that the photon enters the switch right after its state was set. We check the synchronization by running the protocol with Bob's optimal attack, which consists of replacing Bob's detection signal with a continuous electronic signal. The timing is appropriately set when the rate in Alice's verification detector is maximized.   

Note that when performing the protocol with honest parties in our conditions (low dark count rate and low double-pair emission probability), then Alice activates her switch only when Bob actually measures the photon, so she cannot measure any photon in her verification detector. This is expected as we tend to minimize the probability of sanctioning an honest Bob, in order to verify the correctness condition. However, this questions the point of using such an optical switch and fast electronics, just to send void on Alice's verification detector. Physically speaking, this seems equivalent to using the exact same setup with no switch, and send all photons to Bob's verification setup. However, we cannot assume Bob to be honest, even when he is. Therefore, it is of major importance that Alice checks that her state actually is projected on the void, in a cryptographic context. 

\subsection{Reflectivities}\label{reflectivities}

\noindent Now we give a recipe to tune the reflectivities in the experiment, to measure them, and we compare these measurements to our theoretical predictions. 

When parties are honest, Bob first sets $z=1/2$ by blocking Alice's signal, and equalizing the detection rates in $D_{V_1}$ and $D_{V_2}$. This later ensures an optimized interference, and therefore the correctness condition. Then he can tune $y$ such that the detection rate in $D_B$ equals twice the total rate in $D_{V_1}$ and $D_{V_2}$, which should ensure the fairness condition. Alice then tunes $x$ in order to optimize the interference visibility, which should complete the setting of reflectivities. If $v$ is significantly lower than $1$, Alice and Bob might have to perform some mild adjustments on $x$ and $y$ in order to maximize the fairness and correctness.

\begin{figure}[htbp]
	\begin{center}
		\includegraphics[width=110mm]{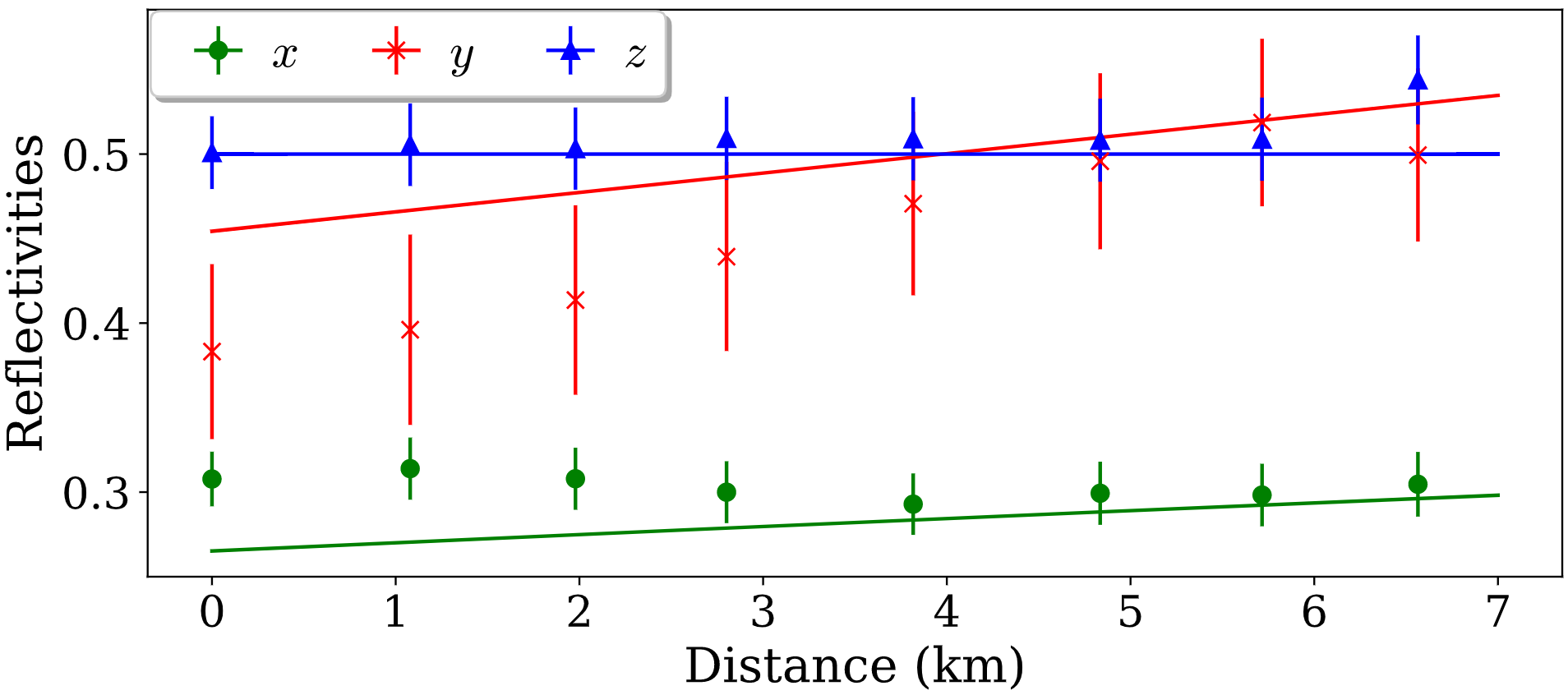}
		\caption{Reflectivities measured in protocols with honest parties, for different communications distances simulated with VOAs. The lines show the prediction from Eqs.~(\ref{eq:xtheo}) to (\ref{eq:ztheo}), with efficiencies given in Table~\ref{tab:Losses}, and with some additional factors $e^{-0.02L}$ induced by VOAs. The error bars are mainly due to error propagation on the efficiencies.}
		\label{fig:reflectivities}
	\end{center}
\end{figure}

After performing a protocol with reflectivities $x,y,z$ we can evaluate them by measuring some specific probabilities. We now give the recipe of this procedure, the results of which are shown in Fig.~\ref{fig:reflectivities}, for protocols with honest parties, and with VOAs simulating different communication distances. First, we force the switch in state $s=1$ and measure the detection probability in detector $D_A$. We can then extract $x$ from the following expression:
\begin{equation}\label{eq:evalX}
    P_{D_A} = x\eta_A^s.
\end{equation}
Then we measure the detection probability in detector $D_B$, and extract $y$ from the expression: 
\begin{equation}\label{eq:evalY}
    P_{D_B} = (1-x)y\eta_B^y.
\end{equation}
Finally, we force the switch in state $s=0$ and block Bob's side of the interferometer, such that the photon does not interfere on his verification BS, and we extract $z$ from one of these expressions: 
\begin{equation}\label{eq:evalz}
\begin{aligned}
    &P_{V_1} = xz\eta_A^{V_1},\\
    &P_{V_2} = x(1-z)\eta_A^{V_2}. 
\end{aligned}
\end{equation}
We see in Fig.~\ref{fig:reflectivities} that the experimentally measured reflectivities can deviate from the theoretical predictions derived from the efficiency values. The most plausible explanation is that we might not perfectly set the expected reflectivities in each protocol run. This could happen if the fairness $\mathcal{F}$ and the correctness $\mathcal{C}$ are scarcely sensitive to reflectivities around the optimal configuration. Also some undetected errors might have occurred when measuring the efficiencies in Table~\ref{tab:Losses}, because of some undetected fluctuations, or if we did not perfectly set the reflectivities $x,y,z$ to trivial values when performing that measurement.

\newpage

\subsection{Measurement of outcome probabilities}

\noindent First let us recall the five mutually incompatible protocol outcomes: 
\begin{itemize}
    \item Alice wins when $(b,v_1,v_2)=(0,1,0)$,
    \item Alice is sanctioned if $(b,v_2) = (0,1)$,
    \item Bob wins when $(b,a) = (1,0)$,
    \item Bob is sanctioned if $(b,a) = (1,1)$,
    \item The protocol aborts if $(b,v_1,v_2)=(0,0,0)$.
\end{itemize}

We evaluate the probabilities of these outcomes by measuring the different detection rates and coincidence rates, provided by a simple function of our time tagger. However, the time tagger does not provide a direct way of measuring the rate of an event excluding some other event. For instance, in order to measure the rate of "Bob wins" event, we need to measure the rate of detection in Bob's detector, that did not occur at the same time as a detection in Alice's verification detector. In logical notation, we need the event $ b \wedge \neg a$. To calculate such event, we use the fact that for any pair of detection events $u,v$, we have $u\wedge \neg v = u\wedge \neg (v\wedge u)$ such that the rate $R_{u\setminus v}$ of that event can be calculated as $R_{u\setminus v} = R_u - R_{uv}$, with $R_u$ the rate of detection $u$ and $R_{uv}$ the rate of simultaneous detections $u$ and $v$. In this way, we can easily deduce the formula for the rates of different outcomes in the protocol, summarized in Table~\ref{tab:outcomes}. 

\begin{table}[htbp]
\begin{tabular}{|c|@\fillcol c @\fillcol|@\fillcol c @\fillcol|@\fillcol c @\fillcol|@\fillcol c @\fillcol|c|c|}
 \hline
  Outcome &\hspace{0.15cm}$a$\hspace{0.15cm} & \hspace{0.15cm}$b$\hspace{0.15cm} & \hspace{0.1cm}$v_1$\hspace{0.1cm} & \hspace{0.1cm}$v_2$\hspace{0.1cm}&\hspace{0.05cm} Logical\hspace{0.1cm}&\hspace{0.1cm} Rate\hspace{0.1cm} \\
  \hline
  Alice wins &\cellcolor{gray}& 0 & 1 & 0 & $\neg b \wedge v_1 \wedge \neg v_2$  & $R_{hV_1} - R_{hV_1V_2} - R_{hBV_1} + R_{hBV_1V_2} $\\
  \hline 
  Bob wins & 0 & 1 & \cellcolor{gray} & \cellcolor{gray} & $b \wedge \neg a$& $R_{hB}-R_{hAB}$\\
  \hline
  Alice is sanctioned  & \cellcolor{gray} & 0 & \cellcolor{gray} & 1 & $\neg b \wedge v_2$& $R_{hV_2} - R_{hBV_2}$\\
  \hline
  Bob sanctioned & 1 & 1 & \cellcolor{gray} &\cellcolor{gray}& $b\wedge a$ & $R_{hAB}$\\
  \hline
  Abort & \cellcolor{gray} & 0 & 0 & 0 & $\neg b \wedge \neg v_1 \wedge \neg v_2$ & $R_h - \{\textrm{\textit{Rates of all other outcomes}}\}$\\
  \hline
\end{tabular}
\caption{Different protocol events, with the corresponding detection outcomes, logical formula and combination of coincidence rates needed to compute the corresponding probability. The rates subscripts correspond to the detectors which simultaneously trigger, $h$ for the heralding, $B$ for Bob's detector, $A$ for Alice's verification detector, $V_1$ and $V_2$ for Bob's verification detectors.}\label{tab:outcomes}
\end{table}

\end{document}